\documentclass[11pt]{article}
\usepackage{lipsum}
\usepackage{inputenc}
\usepackage{nicematrix}
\usepackage{amsthm}
\usepackage{amsmath}
\usepackage[font={scriptsize}]{caption}
\usepackage{multicol}
\usepackage{physics}
\usepackage{esint}
\usepackage{cancel}
\usepackage{titlesec}
\usepackage{stmaryrd}
\usepackage[font=small]{caption}
\usepackage{slashed}
\usepackage{pgfplotstable}
\usepackage{xcolor}
\usepackage{comment}
\usepackage{dsfont}
\usepackage{mathtools}
\usepackage{ulem}
\usepackage{cite}
\usepackage{hyperref}
\hypersetup{
 colorlinks=true,
 linkcolor=black,
 anchorcolor = blue,
 citecolor = blue,
 filecolor = blue,
 urlcolor = blue
}

\usepackage{graphicx}
\usepackage{subfig}
\usepackage{amssymb}
\usepackage{verbatim}
\usepackage{multicol}
\usepackage{bm}
\usepackage[left=1in, right=1in, top=0.92in, bottom=1in]{geometry}
\usepackage{standalone}
\usepackage{tikz}
\usetikzlibrary{decorations.markings}
\usetikzlibrary{shapes.arrows, fadings}
\usetikzlibrary{math} 
\usetikzlibrary{positioning}
\usepackage{float}
\usepackage{pgfplots}

\pgfplotsset{width=10cm, height=7cm, compat=1.9}

\tikzfading[name=fade right,
  left color=transparent!30, right color=transparent]

\definecolor{myred}{HTML}{e41a1c}
\definecolor{myblue}{HTML}{377eb8}
\definecolor{mygreen}{HTML}{4daf4a}

\makeatletter
\newsavebox{\@brx}
\newcommand{\llangle}[1][]{\savebox{\@brx}{\(\m@th{#1\langle}\)}%
  \mathopen{\copy\@brx\kern-0.5\wd\@brx\usebox{\@brx}}}
\newcommand{\rrangle}[1][]{\savebox{\@brx}{\(\m@th{#1\rangle}\)}%
  \mathclose{\copy\@brx\kern-0.5\wd\@brx\usebox{\@brx}}}
\makeatother

\newcommand\flc{\bgroup\markoverwith{\textcolor{blue}{\rule[0.5ex]{2pt}{0.4pt}}}\ULon}

\title{{\bf Non-equilibrium entanglement asymmetry for discrete groups: the example of the XY spin chain }}
\author{Florent Ferro$^1$, Filiberto Ares$^{1, 2}$, and Pasquale Calabrese$^{1, 2, 3}$}
\date{}

\begin{document}

\maketitle
{\small
\vspace{-10mm}  \ \\
{$^{1}$}  SISSA, via Bonomea 265, 34136 Trieste, Italy\\
\medskip
{$^{2}$} INFN Sezione di Trieste, via Bonomea 265, 34136 Trieste, Italy\\[-0.2cm]
\medskip
{$^{3}$}  International Centre for Theoretical Physics (ICTP), Strada Costiera 11, 34151 Trieste, Italy\\
\medskip
}

\begin{abstract}
The entanglement asymmetry is a novel quantity that, using entanglement methods, measures how much a symmetry is broken in a part of an extended quantum system.
So far it has only been used to characterise the breaking of continuous Abelian symmetries. 
In this paper, we extend the concept to cyclic $\mathbb{Z}_N$ groups.  
As an application, we consider the XY spin chain, in which the ground state spontaneously breaks the $\mathbb{Z}_2$ spin parity symmetry in the ferromagnetic phase. 
We thoroughly investigate the non-equilibrium dynamics of this symmetry after a global quantum quench, generalising known results for the standard order parameter.
\end{abstract}

\tableofcontents

\section{Introduction}

In the last few years, the interplay between entanglement and symmetries has become the centre of an intense research activity, yielding up a more refined vision of the behaviour of many-body quantum systems~\cite{lukin, azses, neven, vitale, rath, bcckr-22} and creating a new framework to study entanglement~\cite{lr-14, goldstein, xavier, cgs-18, brc-19, mdgc-20, hc-20, mbc-21, fg-20, mcp-22, chcc-22, dgmnsz-23, cam-23,mdc-19,pbc-21}. For example, novel quantities, such as the symmetry-resolved entanglement entropy~\cite{lr-14, goldstein, xavier}, have been conceived to analyse how entanglement distributes in the symmetry sectors of a theory. On the other hand, it has recently arisen the idea of using entanglement tools to study symmetry breaking. To this end, entanglement asymmetry has been proposed in Ref.~\cite{amc-22} as a measure of how much a symmetry is broken in a part of an extended quantum system. So far, the entanglement asymmetry has been applied to examine the time evolution of an initially broken $U(1)$ symmetry after a quench with a Hamiltonian that preserves it, both in free~\cite{amc-22, amvc-23} and interacting integrable systems~\cite{bkccr-23}. Generally, it is expected that the symmetry is dynamically restored in a subsystem after the quench. However, unexpectedly, the entanglement asymmetry reveals that the symmetry can be restored earlier when it is initially more broken~\cite{amc-22}. This surprising phenomenon can be seen as a quantum version of the very counter-intuitive Mpemba effect: the more a system is initially out of equilibrium, the faster it relaxes. Entanglement asymmetry has been also employed in Ref.~\cite{amvc-23} to analyse the special case in which symmetry is not restored and the subsystem relaxes to a non-Abelian Generalised Gibbs ensemble.

The previous works restrict to the continuous $U(1)$ group. The goal of the present manuscript is to extend the study of the entanglement asymmetry to the finite, discrete cyclic $\mathbb{Z}_N$ groups. In particular, we consider the $\mathbb{Z}_2$ spin flip parity symmetry of the XY spin chain which, in the thermodynamic limit, is spontaneously broken by the ground state in the ferromagnetic phase~\cite{sachdev}. Usually this symmetry breaking is detected using the one-site longitudinal magnetisation as order parameter. This order parameter presents the inconvenience that, despite a non-zero value indicating that the symmetry is broken, the converse is not always true~\cite{bcp-23}. On the contrary, the entanglement asymmetry unambiguously determines if the symmetry is or not respected. Moreover, although for a single spin the order parameter can be used to measure symmetry breaking, for larger subsystems there are longer range correlations that can break the symmetry and are only taken into account by the entanglement asymmetry. Using it, we study the non-equilibrium time evolution of the $\mathbb{Z}_2$ spin parity symmetry after a global quantum quench in the XY spin chain Hamiltonian, extending the study performed in Refs.~\cite{cef-11, cef-12-1, cef-12-2} employing the
order parameter. Global quenches in this system have been studied from many perspectives, see e.g.~\cite{bmd-70, ir-00, sps-04,cc-05, fc-08, rsms-09, ir-10, eef-12, fe-13, fe-16, gec-18,gde-22,bkc-14,kbc-14,pb-22}. We distinguish between chains with open and periodic boundary conditions, since the approach to calculate the entanglement asymmetry and the results obtained differ: while in the periodic case this $\mathbb{Z}_2$ symmetry is always restored, the same does not always occur in the open chain due the presence of boundary modes. Moreover, in the periodic case, by comparing with the analytic results obtained in Refs.~\cite{cef-11, cef-12-1} for the order parameter in the thermodynamic limit, we conjecture an effective expression that describes the evolution of the entanglement asymmetry after the quench.

The paper is organised as follows: in Sec.~\ref{sec:ent_asymm}, we define the entanglement asymmetry, we review its main known features for the $U(1)$ group and we extend it to cyclic $\mathbb{Z}_N$ groups, describing some of its principal properties in this case. In Sec.~\ref{sec:xy_chain}, we introduce the XY spin chain and we present the behaviour of the entanglement asymmetry of the spin parity in the ground state. 
Sec.~\ref{sec:obc} is devoted to calculate and discuss the entanglement asymmetry after a quantum quench of the XY spin chain with OBC, while in Sec.~\ref{sec:pbc} we carry out a similar analysis but taking PBC. In Sec.~\ref{sec:conclusions}, we draw our conclusions. The manuscript also includes several appendices that describe in detail those points of the main text that are more technical.

\section{Entanglement asymmetry}\label{sec:ent_asymm}

Let us consider an extended quantum system that can be divided into two spatial parts $A$ and $B$. If we assume that the total system is in a pure state $\ket{\Psi}$, then the state of one of the regions, for example $A$, is described by the reduced density matrix $\rho_A = \Tr_B (\ket \Psi \bra \Psi)$, which is generally a mixed state. 
We further take a local charge operator $Q$ given by the sum of the charge in $A$ and $B$, $Q=Q_A+Q_B$. We assume that $Q$ generates an Abelian group with elements of the form $e^{i\alpha Q}$ where $\alpha$ is a real number. This includes the unitary group $U(1)$ or the cyclic groups $\mathbb{Z}_N$. 

It is well-known that, when $\ket{\Psi}$ respects the symmetry associated to $Q$, i.e. it is an eigenstate of $Q$, the reduced density matrix of any subsystem $A$ displays a block-diagonal structure in the eigenbasis of $Q_A$, and the blocks correspond to the sectors of a given charge. In this case, the entanglement entropy of subsystem $A$ can be decomposed into the contributions of each charge sector, known as symmetry-resolved entanglement entropies \cite{lr-14, goldstein, xavier}.

In the case of symmetry-breaking states, entanglement entropy can be used to quantify how much the symmetry generated by $Q$ is broken in a subsystem. In fact, if the state $\ket{\Psi}$ locally breaks the symmetry in the region $A$, then $\rho_A$ does not commute with $Q_A$. This means that, in the eigenbasis of $Q_A$, $\rho_A$ presents non-zero entries outside the sectors of fixed charged $\rho_{q_j}$,

\NiceMatrixOptions{cell-space-limits=2mm, extra-margin=5pt}
\[\rho_A = \begin{pNiceMatrix}[columns-width=0.4cm]
 \Block[draw,fill=blue!15,rounded-corners]{1-1}{\rho_{q_1}} & * & * & * \\
  *& \Block[draw,fill=blue!15,rounded-corners]{2-2}{\rho_{q_2}} &  & * \\
  *&  &  & * \\
  *& * & * & \Block[draw,fill=blue!15,rounded-corners]{1-1}{\rho_{q_3}}  \\
\end{pNiceMatrix},\]
here represented by $*$. From this state, it is always possible to define another density matrix $\rho_{A, Q}$ that respects the symmetry, i.e. $[\rho_{A, Q}, Q_A]=0$. If we take the projection of $\rho_A$ on the different charge sectors of $Q_A$ and we sum all them, we obtain a block diagonal matrix

\NiceMatrixOptions{code-for-first-row = \scriptstyle, cell-space-limits=2mm, extra-margin=5pt}
\[\rho_{A, Q} = \sum_{q_j}\Pi_{q_j} \rho_A \Pi_{q_j} = \begin{pNiceMatrix}[columns-width=0.4cm]
 \Block[draw,fill=blue!15,rounded-corners]{1-1}{\rho_{q_1}} &  &  &  \\
  & \Block[draw,fill=blue!15,rounded-corners]{2-2}{\rho_{q_2}} &  &  \\
  &  &  &  \\
  &  &  & \Block[draw,fill=blue!15,rounded-corners]{1-1}{\rho_{q_3}}  \\
\end{pNiceMatrix},\]
where $\Pi_{q_j}$ is the projector on the subspace of eigenvalue $q_j$. The density matrix $\rho_{A, Q}$ can be physically interpreted as the result of performing a non-selective measurement of $Q_A$ in $\rho_A$. From $\rho_A$ and its projection $\rho_{A, Q}$, we can define the entanglement asymmetry~\cite{amc-22}
\begin{equation}\label{eq:ent_asymm}
    \Delta S_A= S(\rho_{A, Q})-S(\rho_A),
\end{equation}
in terms of their von Neumann entropies
\begin{equation}
    S(\rho) = -\Tr(\rho\log\rho).
\end{equation}
The entanglement asymmetry measures how much the symmetry generated by $Q$ is broken in the susbsystem $A$. In fact, $\Delta S_A$ satisfies two crucial properties: it is always non-negative, $\Delta S_A\geq 0$, and it vanishes if and only if $[\rho_A, Q_A]=0$~\cite{mhms-22}. These properties are satisfied whether the total system is in a pure or mixed state. Therefore, the entanglement asymmetry can also be applied to the study of symmetry breaking in mixed states and, in particular, to a system at a certain temperature.

Since the direct calculation of the von Neumann entropy is normally a difficult task, the usual strategy~\cite{hlw-94, cc-04} is to replace it in Eq.~\eqref{eq:ent_asymm} by the Rényi entropies
\begin{equation}
S^{(n)}(\rho)=\frac{1}{1-n}\log \Tr(\rho^n).
\end{equation}
Then we define the Rényi entanglement asymmetry as
\begin{equation}\label{eq:def_ren_asym}
\Delta S_A^{(n)}=S^{(n)}(\rho_{A, Q})-S^{(n)}(\rho_A).
\end{equation}
Note that, if we take in this expression the limit $n\to 1$, we 
recover Eq.~\eqref{eq:ent_asymm}. Moreover, Rényi entanglement asymmetries are also non-negative quantities, $\Delta S_A^{(n)}\geq 0$, and they cancel if and only if the symmetry is respected in subsystem $A$, i.e. $[\rho_A, Q_A]=0$~\cite{hms-23}. It is also important to remark that the Rényi entanglement asymmetries for integer $n\geq 2$ are experimentally measurable in ion traps through randomised measurements~\cite{vermersch, brydges-19, huang, elben,mp-exp}.

\subsection{Review of entanglement asymmetry for the $U(1)$ group}\label{sec:tilted_ferro}

In the literature, the entanglement asymmetry has only been analysed for the Abelian unitary group $U(1)$. In this case, the paradigmatic example to understand how it works is the tilted ferromagnetic state,
\begin{equation}\label{eq:tilted_ferro}
    \ket \theta = e^{i\frac \theta 2  \sum_{j} \sigma^y_j} \ket{\uparrow \uparrow \uparrow \cdots}.
\end{equation}
For $\theta \neq 0$, this state breaks the $U(1)$ symmetry corresponding to rotations around the $z$ axis, which are generated by the transverse magnetisation $Q = \frac 1 2 \sum_j \sigma^z_j$. It has been shown that, if the subsystem $A$ is an interval of contiguous spins of length $\ell$, the entanglement asymmetry behaves as \cite{amc-22}
\begin{equation}\label{eq:ent_asymm_tilted_ferro}
    \Delta S_A^{(n)} = \frac 1 2 \log \ell + \frac 1 2 \log \frac{\pi n^{\frac{1}{n-1}}\sin^2 \theta }{2} + \mathcal{O}(\ell^{-1})
\end{equation}
for large $\ell$. To better understand this result, let us focus on the $n=2$ Rényi asymmetry $\Delta S_A^{(2)}$. Using the Fourier representation of the projectors $\Pi_{q_j}$, the projected matrix $\rho_{A, Q}$ can be written as
\begin{equation}\label{eq:fourier_rhoAQ}
    \rho_{A, Q} = \int_{0}^{2\pi} \frac{d\alpha}{2\pi} e^{i\alpha Q_A} \rho_A e^{-i\alpha Q_A}.
\end{equation}
Since $\ket{\theta}$ is a separable state, $\rho_A=\ket{\theta, \ell}\bra{\theta,\ell}$, and $\ket{\theta,\ell}$ is the restriction of $\ket{\theta}$ to $\ell$ contiguous spins.
We can see Eq. \eqref{eq:fourier_rhoAQ} as the state of $A$ after applying a random element of the $U(1)$ group to $\rho_A$ with a uniform probability density. Consequently, $\Delta S_A$ can be interpreted as the quantum information loss about the subsystem $A$ after such an operation. Despite the infinite number of elements in the $U(1)$ group, this information loss is finite because two states $\ket{\theta, \ell}$ and $e^{i\alpha Q_A}\ket{\theta, \ell}$ are not orthogonal for $\alpha\neq 0$ and finite $\ell$ but have a non-zero overlap. In fact, the $n=2$ Rényi entanglement asymmetry can be written as
\begin{equation}\label{eq:average}
    \Delta S_A^{(2)} = -\log \left(\int_{0}^{2\pi} \frac{d\alpha}{2\pi}|\bra{\theta, \ell}e^{i\alpha Q_A}\ket{\theta, \ell}|^2 \right).
\end{equation}
This integral is the average overlap between the states $\ket{\theta, \ell}$ and $e^{i\alpha Q_A}\ket{\theta, \ell}$ with respect to $\alpha$. Hence, according to Eq.~\eqref{eq:average}, $\Delta S_A^{(2)}$ grows as the overlap between $\ket{\theta,\ell}$ and $e^{i\alpha Q_A}\ket{\theta,\ell}$ decreases and, consequently, the maximal entanglement asymmetry is obtained when $\ket{\theta, \ell}$ and $e^{i\alpha Q_A}\ket{\theta,\ell}$ are orthogonal for all $\alpha\in(0,2\pi)$. A simple calculation gives that
\begin{equation}\label{eq:overlap}
 |\bra{\theta,\ell}e^{i\alpha Q_A}\ket{\theta,\ell}|^2=\left(\cos^2(\alpha/2)+\cos^2\theta\sin^2(\alpha/2)\right)^\ell,
\end{equation}
which, for $\alpha,\theta\in (0,2\pi)$, only vanishes in the limit $\ell\to \infty$. Therefore, for a given value of the tilting angle $\theta$, the entanglement asymmetry of the $U(1)$ group generated by the transverse magnetisation is unbounded. Moreover, using Eq.~\eqref{eq:overlap}, we obtain that the average overlap between $\ket{\theta,\ell}$ and $e^{i\alpha Q_A}\ket{\theta,\ell}$ decreases as $1/\sqrt{\ell}$ for large $\ell$, giving rise to the logarithmic divergence of $\Delta S_A^{(2)}$ in Eq.~\eqref{eq:ent_asymm_tilted_ferro}.

\subsection{Entanglement asymmetry for cyclic groups $\mathbb{Z}_N$}\label{sec:ent_asymm_ZN}

Let us now consider the finite cyclic group $\mathbb{Z}_N$ of rotations around the $z$ axis by angles multiple of $2\pi/N$, which is a subgroup of the $U(1)$ symmetry generated by the transverse magnetisation $Q$. The density matrix $\rho_{A, Q_N}$ of this group is obtained by projecting on the different charge sectors of the transverse magnetisation $Q$ modulo $N$. Using the Fourier representation of the projectors, it can be written as
\begin{equation}\label{eq:rhoaQ_ZN}
    \rho_{A, Q_N} = \frac{1}{N}\sum_{j=1}^{N} e^{i\alpha_j Q_A} \rho_A e^{-i\alpha_j Q_A}, \quad \alpha_j = \frac{2\pi j}{N}. 
\end{equation}

\begin{figure}[t]%
    \centering
    \includegraphics[width=0.49\textwidth]{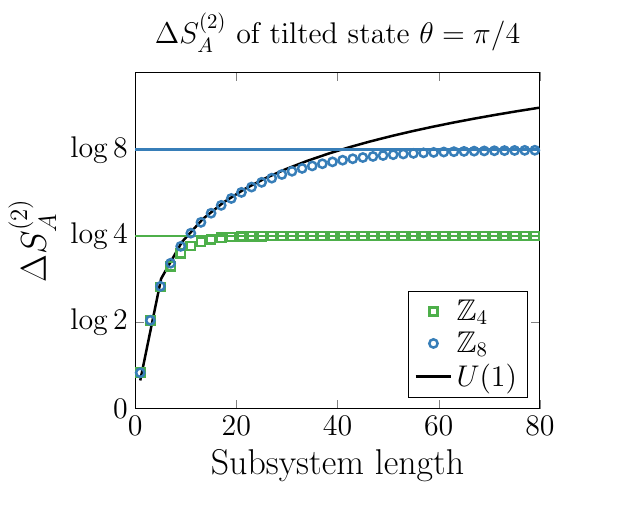}
    \includegraphics[width=0.49\textwidth]{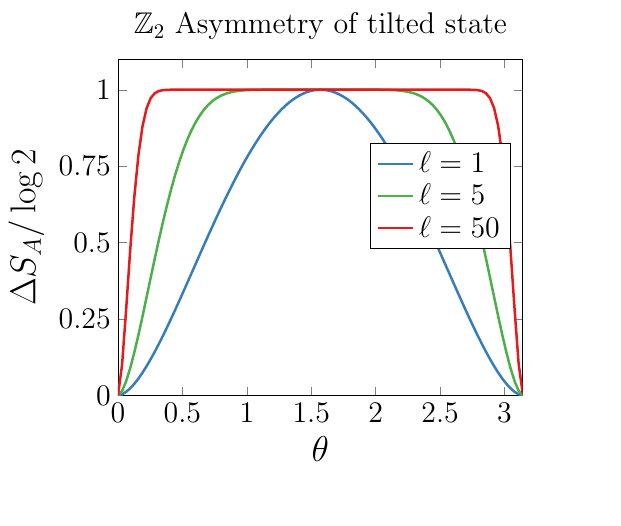} 
    \caption{Left panel: the symbols are the entanglement asymmetry for different subgroups $\mathbb{Z}_N$ of the $U(1)$ group of rotations around the $z$ in the tilted ferromagnetic state~\eqref{eq:tilted_ferro} with $\theta=\pi/4$. We represent them as a function of the subsystem length. For large subsystems they tend to $\log N$. The solid line is  the asymptotic entanglement asymmetry for the  $U(1)$ group, cf.  Eq.~\eqref{eq:ent_asymm_tilted_ferro}, which diverges logarithmically with $\ell$. Right panel: the exact analytic expression~\eqref{eq:ent_asymm_tilted_ferro_Z2} for the subgroup $\mathbb{Z}_2$ as a function of the tilting angle $\theta$ of the ferromagnetic state and different subsystem sizes.} 
    \label{fig:example}%
\end{figure}

\begin{figure}[t]
\centering
   \includegraphics[width=0.45\textwidth]{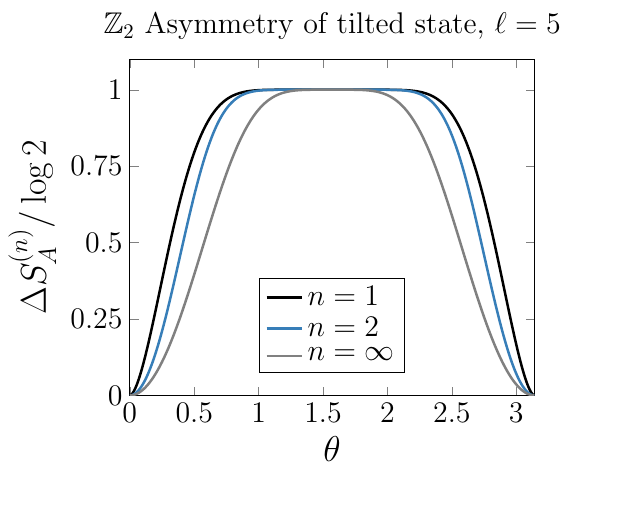} 
    \includegraphics[width=0.45\textwidth]{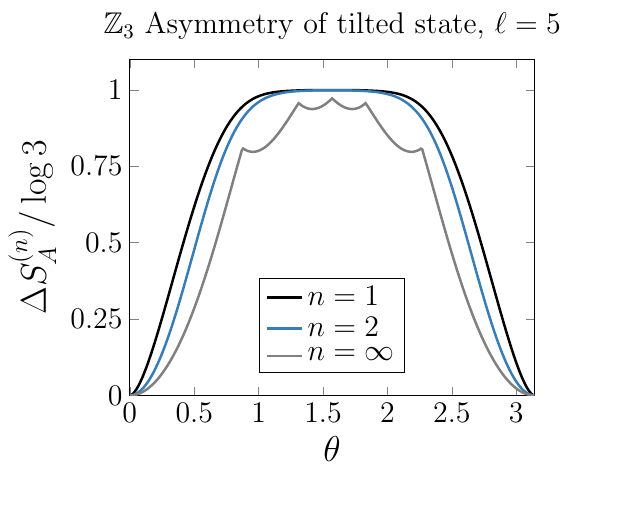} 
    \caption{R\'enyi entanglement asymmetry in the tilted ferromagnetic state for the subgroups $\mathbb{Z}_2$ (left panel) and $\mathbb{Z}_3$ (right panel) of the $U(1)$ group of rotations around the $z$ axis. We take different R\'enyi indices $n$ and subsystem length $\ell=5$.}%
    \label{fig:monotonicity}
\end{figure}

This matrix can be interpreted either as the state of subsystem $A$ after a non-selective measurement of the transverse magnetisation $Q$ modulo $N$ or after randomly applying to $\rho_A$ one of the elements of the $\mathbb{Z}_N$ group with equal probabilities. The second interpretation  gives the intuition that the maximal information loss after such an operation is $\log N$ since $N$ is the dimension of the group and one can always recover all the information on the state of $A$ by learning which element of the group has been applied. In the left panel of Fig.~\ref{fig:example}, we study the entanglement asymmetry of the tilted ferromagnetic state $\ket{\theta}$ as a function of the subsystem length $\ell$ for different subgroups $\mathbb{Z}_N$ of the $U(1)$ group of rotations around the $z$ axis seen before. 
For small subsystem size, the entanglement asymmetry presents the same behaviour as in the $U(1)$ case. However, for large $\ell$, it precisely saturates to $\log N$. This saturation occurs at larger values of $\ell$ as the dimension of the subgroup increases, recovering the result of Eq.~\eqref{eq:ent_asymm_tilted_ferro} for the $U(1)$ group in the limit $N\to \infty$. In particular, for $N=2$, we have
\begin{equation}\label{eq:ent_asymm_tilted_ferro_Z2}
    \Delta S_A^{(n)} = H^{(n)}(\cos^\ell \theta),
\end{equation}
where 
\begin{equation}\label{eq:hncircle}
    H^{(n)}(x) = \frac{1}{1-n}\log \left[\left(\frac{1+x}{2}\right)^n + \left(\frac{1-x}{2}\right)^n\right]
\end{equation}
are the functions for the Rényi-$n$ entropies of one bit of information. Their limit $n\to 1$ reads
\begin{equation}\label{eq:hnto1}
    H^{(1)} (x) := \lim_{n\to 1} H^{(n)}(x) = -\frac{1+x}{2}\log(\frac{1+x}{2})
    -  \frac{1-x}{2}\log(\frac{1-x}{2}).
\end{equation}
In the right panel of Fig.~\ref{fig:example}, we plot Eq.~\eqref{eq:ent_asymm_tilted_ferro_Z2} (i.e. for $N=2$) as a function of the tilting angle $\theta$ for $n=1$ and different subsystem sizes. We can see that $\Delta S_A$ grows monotonically with $\theta$ until $\theta=\pi/2$, value at which the $U(1)$ symmetry is maximally broken since all the spins point in the $x$ direction. Observe that, as the subsystem size increases, the asymmetry saturates faster to its maximum value, $\log 2$, when the tilted angle $\theta$ turns on. 
In Fig.~\ref{fig:monotonicity}, we study the R\'enyi entanglement asymmetry for the subgroups $\mathbb{Z}_2$ (left panel) and $\mathbb{Z}_3$ (right panel) as a 
function of the tilting angle $\theta$ for different R\'enyi indices $n$ and a fixed subsystem length. As we can see, for the subgroup $\mathbb{Z}_2$, $\Delta S_A^{(n)}$ is monotonic in $\theta$ for $n\to\infty$. On the other hand, for $\mathbb{Z}_3$, it is 
non monotonic in $\theta$ when $n\to\infty$, presenting the same peaked behaviour as in the $U(1)$ case~\cite{amc-22}. This is expected since, as we show in the left panel of Fig.~\ref{fig:example}, $\Delta S_A^{(n)}$ for the subgroups $\mathbb{Z}_N$ tends to the $U(1)$ asymmetry in the limit $N\to \infty$.

We can generally prove that, for $\mathbb{Z}_N$ groups, the 
entanglement asymmetry $\Delta S_A$ is bounded by the dimension of 
the group. This is actually a consequence of the following inequality for the von Neumann entropy. If we consider a set of density matrices $\{\rho_j\}$ and of positive coefficients $\{\lambda_j\}$ with $\sum_j \lambda_j=1$, then~\cite{nielsen}
\begin{equation}\label{eq:ent_ineq} 
S\left(\sum_j \lambda_j \rho_j\right)\leq \sum_j \lambda_j S(\rho_j) -\sum_j \lambda_j \log\lambda_j.
\end{equation}
Therefore, if in the definition~\eqref{eq:ent_asymm} of the entanglement asymmetry we take into account that for a $\mathbb{Z}_N$ group the projected density matrix $\rho_{A, Q_N}$
can be written in the form of Eq.~\eqref{eq:rhoaQ_ZN}, and we apply the previous inequality identifying $\lambda_j=1/N$ and $\rho_j=e^{i\alpha_j Q_A} \rho_A e^{-i\alpha_j Q_A}$, then
we can easily conclude that 
\begin{equation}
\Delta S_A\leq \log N.
\end{equation}
The equality in Eq.~\eqref{eq:ent_ineq} is attained if and only if the states $\rho_j$ have support in orthogonal 
subspaces. This implies that the entanglement asymmetry for a $\mathbb{Z}_N$ group saturates to
its maximal value, $\Delta S_A=\log N$, if and only if all the states $e^{i\alpha_j Q_A}\rho_A e^{-i\alpha_j Q_A}$ 
have orthogonal support. Therefore, the $\mathbb{Z}_N$ symmetry is maximally broken if the action of any element of the group, except the identity, maps $\rho_A$ to different orthogonal subspaces. In the tilted ferromagnetic state, this is equivalent to requiring that $\ket{\theta,\ell}$ and $e^{i\alpha_j Q_A}\ket{\theta,\ell}$ are orthogonal, similarly to the discussion done in Sec.~\ref{sec:tilted_ferro} for the $U(1)$ case, but now the average~\eqref{eq:average} is over a finite number of states and, therefore, is bounded. In Appendix \ref{appendix:bound},  we prove that these properties are also satisfied by the Rényi entanglement asymmetry~\eqref{eq:def_ren_asym} for any index $n$. Interestingly, the equality in Eq.~\eqref{eq:ent_ineq} corresponds to a particular instance of Holevo theorem~\cite{nielsen}, a result of quantum information theory that states that the maximum information that one can extract from a mixed state such as $\rho_{A, Q_N}$ is given by the difference between $S(\sum_j\lambda_j\rho_j)$ and $\sum_j\lambda_j S(\rho_j)$, which in our case is precisely $\log N$, as we already announced.

\section{The XY spin chain}\label{sec:xy_chain}

In the rest of the paper, we study the non-equilibrium dynamics of a $\mathbb{Z}_2$ broken symmetry in the XY spin chain. This section is devoted to introducing this model and discuss some of its general features that will be useful afterwards. The Hamiltonian of the XY spin chain reads
\begin{equation}\label{eq:HXY}
    H = -\sum_{j=1}^{L-1}\left[ \frac{1+\gamma}{4}\sigma^x_j \sigma^x_{j+1} + \frac{1-\gamma}{4}\sigma^y_j \sigma^y_{j+1}\right] - \frac{1}{2}\sum_{j=1}^{L}  h\sigma^z_j\,\; (+ H_b),
\end{equation}
where
\begin{equation}
    H_b = - \left[ \frac{(1+\gamma)}{4}\sigma^x_L \sigma^x_{1} + \frac{(1-\gamma)}{4}\sigma^y_L \sigma^y_{1}  \right]
\end{equation}
is a boundary term closing the chain with Periodic Boundary Conditions (PBC). If this term is not present, then the chain has Open Boundary Conditions (OBC). Observe that the Hamiltonian contains nearest neighbour interactions between the $x$ and $y$ components of the spins. The parameter $\gamma>0$ tunes the anisotropy between the couplings in these two directions. There is also an external magnetic field of intensity $h>0$ in the $z$ direction.

The reason why we are interested in this system is that it exhibits a $\mathbb{Z}_2$ symmetry corresponding to rotations of $\pi$ rad around the $z$ axis, a subgroup of the $U(1)$ symmetry discussed in Sec.~\ref{sec:tilted_ferro} and in Refs.~\cite{amc-22, amvc-23}. It is associated to the parity operator
 \begin{equation}\label{eq:parity_op}
 P = \prod_{j=1}^L \sigma^z_j = (-i)^L e^{i\pi Q},\quad 
 \text{where}\,\,  Q=\frac{1}{2}\sum_{j=1}^L\sigma_j^z,
 \end{equation}
  which maps the spins $\sigma_j^{x, y}$ to 
 $-\sigma_j^{x,y}$ leaving $\sigma_j^z$ invariant. This symmetry is spontaneously broken by the ground state of Eq.~\eqref{eq:HXY} in the thermodynamic limit $L\to\infty$ when $h<1$
 (ferromagnetic phase) while, for $h>1$ (paramagnetic phase), the symmetry is respected. Both phases are separated by the critical line
 $h=1$. 

 These two phases can be distinguished using the one-site longitudinal magnetisation $\langle \sigma_j^x\rangle$ as order parameter. In fact, for $h<1$, we have that $\langle \sigma_j^x \rangle \neq 0$ while, for $h>1$,  $\langle \sigma_j^x\rangle = 0$. Since a non-zero magnetisation in the $x$ direction obviously breaks the $\mathbb{Z}_2$ symmetry of $\pi$ rad rotations about the $z$ axis, then the condition $\langle \sigma_j^x\rangle\neq 0$ implies that such symmetry is broken in the ground state.

\begin{figure}[t]
    \centering
    \includegraphics[width=6cm]{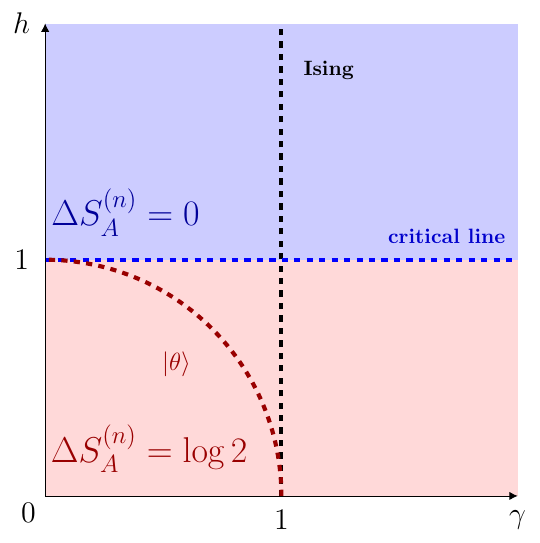}
    \caption{$(h,\gamma)$-parameter plane of the XY spin chain~\eqref{eq:HXY}. For $\gamma>0$, the system is critical (zero mass gap) along the line $h=1$, which separates the ferromagnetic ($h<1$) and the paramagnetic ($h>1$) phases. The 
    line $\gamma=1$ corresponds to the quantum Ising chain.
    In the thermodynamic limit, the ground state spontaneously breaks $\mathbb{Z}_2$ spin parity symmetry in the ferromagnetic phase and the corresponding entanglement asymmetry quickly saturates to $\log 2$ in all the region when the subsystem is large enough.  
    A special line is $h^2+\gamma^2=1$, in which the ground state is the tilted ferromagnetic state $\ket{\theta}$ analysed in Secs.~\ref{sec:tilted_ferro} and \ref{sec:ent_asymm_ZN}.}
    \label{fig:xyphasediagram}
\end{figure}

 We can also analyse this spontaneous breaking of the $\mathbb{Z}_2$ parity symmetry with the Rényi entanglement asymmetry~\eqref{eq:def_ren_asym}. In Fig.~\ref{fig:xyphasediagram}, we represent the $(h, \gamma)$-parameter plane  of the XY spin chain, indicating the large subsystem value of the asymmetry in each 
 phase. We find that, in the ferromagnetic phase $h<1$, $\Delta S_A^{(n)}$ quickly saturates for any Rényi index $n$ to its maximal value, $\log 2$, as the length of the subsystem $\ell$ increases. On the contrary, in the paramagnetic region $h>1$, the asymmetry vanishes, signaling that the ground state respects the symmetry.
 Therefore, for infinitely large subsystems $\Delta S_A^{(n)}$ acts somehow as a topological indicator of symmetry-breaking. It is also worth mentioning that the tilted ferromagnetic state $\ket{\theta}$, for which we obtained its exact $\mathbb{Z}_2$ entanglement asymmetry in Sec.~\ref{sec:ent_asymm_ZN}, is the ground state of the XY spin chain along the circle $\gamma^2+h^2=1$~\cite{ktm-82, ms-85}.

 In this work, we are interested in studying the time evolution of this $\mathbb{Z}_2$ symmetry in a global quantum quench from the ground state of the Hamiltonian with couplings $(h_0, \gamma_0)$ in the ferromagnetic phase, i.e. $h_0<1$, to another XY Hamiltonian with different couplings $(h_f, \gamma_f)$. One has $[H, P]=0$ for any post-quench XY Hamiltonian $H$. Hence, after this quench, the state that describes a subsystem $A$ is expected to relax to a Generalised Gibbs Ensemble (GGE) that respects the $\mathbb{Z}_2$ symmetry~\cite{cef-12-1, cef-12-2, fe-13}. This symmetry restoration has actually been studied using the order parameter $\langle \sigma_j^x\rangle$ in Refs.~\cite{cef-11, cef-12-1}. It was found to exhibit an exponential decay in time, both for quenches to the ferromagnetic and paramagnetic phases. In our analysis, we distinguish two cases. In Sec.~\ref{sec:obc}, we consider the entanglement asymmetry  of an interval attached to one of the extremes of an open chain, situation that in the thermodynamic limit 
$L\to \infty$ corresponds to a semi-infinite chain. In Sec.~\ref{sec:pbc}, we examine the entanglement 
asymmetry in the thermodynamic limit of a chain with periodic boundary conditions. In each case, we apply a 
different approach and we find that, in some situations, the evolution of the asymmetry after the quench differs. 

\subsection{Jordan-Wigner transformation}

One of the most important features of the XY spin chain~\eqref{eq:HXY}
is that it is free, i.e. it can be mapped to a Hamiltonian describing non-interacting spinless fermions~\cite{lieb-61}. This can be done by means of the Jordan-Wigner transformation
\begin{equation}\label{eq:jordanwigner}
    c_j = \left[\prod_{l=1}^{j-1}\sigma^z_l\right] \frac{\sigma^x_j + i\sigma^y_j}{2}, \quad c_j^\dagger = \left[\prod_{l=1}^{j-1}\sigma^z_l\right] \frac{\sigma^x_j - i\sigma^y_j}{2}, \quad \sigma_j^z=1-2c_j^\dagger c_j,
\end{equation}
where $c_j$ and $c_j^\dagger$ are creation and annihilation fermionic operators, satisfying the anticommutation relations
\begin{equation}
    \{c_j, c_{j'}\} = \{c_j^\dagger, c_{j'}^\dagger\} = 0\text{ and } \{c_j^\dagger, c_{j'}\} = \delta_{jj'}.
\end{equation}
Under this mapping, the parity operator~\eqref{eq:parity_op} becomes
\begin{equation}\label{parityfermions}
    P = \prod_{j=1}^L (1-2c_j^\dagger c_j) = e^{i\pi \mathcal N},\quad \text{ where } \mathcal N = \sum_{j=1}^L c_j^\dagger c_j,
\end{equation} 
which corresponds to the parity of the number of fermions $\mathcal N$. The Hamiltonian~\eqref{eq:HXY} is mapped to the Kitaev fermionic chain
\begin{equation}\label{eq:kitaev_chain}
H = -\frac{1}{2}\left[\sum_{j=1}^{L-1}(c_j^\dagger c_{j+1}+c_{j+1}^\dagger c_j)+\gamma (c_j^\dagger c_{j+1}^\dagger+c_{j+1}c_j)+ \sum_{j=1}^{L} h(1-2 c_j^\dagger c_j)\right] (+ H_b),
\end{equation}
which is local in terms of the fermionic operators in spite of the non-locality of the Jordan-Wigner transformation. Observe that the transverse magnetic field  $h$ in Eq.~\eqref{eq:HXY} translates into a chemical potential and the anisotropy parameter $\gamma$ controls the superconducting pairing terms $c_j^\dagger c_{j+1}^\dagger +h.c.$. Importantly, these pairing terms break the fermion number symmetry, but not their parity since they create/destroy particles by pairs.

The boundary term $H_b$ reads after the Jordan-Wigner transformation as
\begin{equation}\label{eq:boundary}
    H_b = \frac{1}{2}P\left[(c_L^\dagger c_{1}+c_{1}^\dagger c_L)+\gamma (c_L^\dagger c_{1}^\dagger+c_{1}c_L)\right].
\end{equation}
Observe that it includes the parity operator $P$ and, therefore, depending on the parity sector, yields periodic or antiperiodic boundary conditions in the Kitaev chain.

\section{Entanglement asymmetry in an open chain}\label{sec:obc}

For OBC, there is no boundary term $H_b$ in the XY spin chain Hamiltonian~\eqref{eq:HXY}, which is a quadratic form in terms of the fermionic operators $c_j$ and $c_j^\dagger$, as we have seen in Eq.~\eqref{eq:kitaev_chain}.
In order to simplify the calculations and to better understand some physical results later, it is very useful to introduce the Majorana fermionic operators~\cite{vlrk-03}
\begin{equation}
    \check a_{2j-1} = c_j^\dagger + c_j \text{ and } \check a_{2j} = i(c_j^\dagger - c_j),
\end{equation}
 by doubling the sites of the lattice. These operators are real valued, $\check a_m^\dagger = \check a_m$, and satisfy the anticommutation relation
\begin{equation}
    \{ \check a_m, \check a_{m'}\} = 2\delta_{mm'}.
\end{equation}

Since the Hamiltonian~\eqref{eq:kitaev_chain} is quadratic, following Ref.~\cite{lieb-61}, it can be diagonalised by performing a Bogoliubov rotation to a new set of fermionic operators $\eta_k$ and $\eta_k^\dagger$. If we organise the Majorana $\check a_m$ and Bogoliubov operators $\eta_k$, $\eta_k^\dagger$ in single vectors,
\begin{equation}
\boldsymbol{\eta} =\begin{pNiceMatrix}
 \eta_{k_0}^\dagger  \\
 \eta_{k_0} \\
 \Vdots \\
 \eta_{k_{L-1}}^\dagger  \\
 \eta_{k_{L-1}}
\end{pNiceMatrix},\quad \boldsymbol{\check a} = \begin{pNiceMatrix}
 \check a_{1}\\
 \check a_{2} \\
 \Vdots \\
 \check a_{2L-1}  \\
 \check a_{2L}
\end{pNiceMatrix},
\end{equation}
then there exists a matrix $\mathcal{R}_{h,\gamma}$ of dimension $2L\times 2L$ that relates $\boldsymbol{\eta}$ and $\boldsymbol{\check{a}}$
\begin{equation}\label{eq:bogoliubov_rot}
    \boldsymbol{\eta} = \mathcal{R}_{h,\gamma} \boldsymbol{\check a}
\end{equation}
such that the Hamiltonian~\eqref{eq:kitaev_chain} is diagonal in terms of the Bogoliubov modes,
\begin{equation}
H=\sum_{j=0}^{L-1} \epsilon_{k_j} \left(\eta_{k_j}^\dagger \eta_{k_j} -\frac{1}{2}\right),
\end{equation}
where 
\begin{equation}\label{eq:disp_rel}
\epsilon_k=\sqrt{(h-\cos k)^2+\gamma^2\sin^2k}
\end{equation}
is the single-particle dispersion relation and the momenta $k_j$ satisfy a particular quantisation condition. In actual applications, 
we will restrict for simplicity to the line $\gamma=1$ (quantum Ising chain). In Appendix~\ref{app:diag_obc}, we report in detail how we obtain the matrix $\mathcal{R}_{h, \gamma}$ in that case.

\begin{figure}[t]
  \centering
  \includegraphics[width=.8\textwidth]{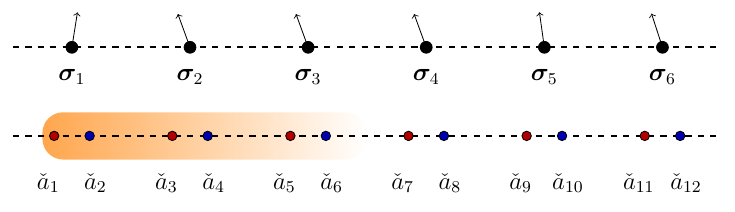}
  \caption{The initial ground state~\eqref{eq:gs_obc} considered to study the entanglement asymmetry of spin parity in OBC contains a boundary Majorana excitation $\check b$ localised at the extreme of the chain indicated above.}
  \label{fig:boundary_mode}
\end{figure}

Therefore, the eigenstates of $H$ are the different configurations $\mathbb{K}$ of occupied Bogoliubov modes
\begin{equation}
\ket{\mathbb{K}}=\prod_{k_j\in\mathbb{K}} \eta_{k_j}^\dagger \ket{\emptyset},
\end{equation}
where $\ket{\emptyset}$ is the Bogoliubov vacuum defined as $\eta_k\ket{\emptyset}=0$ for all $k$ satisfying the quantisation condition. In particular, for $h<1$, $\eta_{k_0}^\dagger\ket{\emptyset}$ is a boundary bound state, see also Appendix~\ref{app:diag_obc}. It can be re-expressed as a Majorana fermionic excitation $\check b \ket{\emptyset}$ localised at one of the edges of the chain, as we illustrate in Fig.~\ref{fig:boundary_mode}, with
\begin{equation}\label{eq:boundary_mode}
 \check b=\eta_{k_0}+\eta_{k_0}^\dagger=\sum_{m=1}^{2L} \beta_m\check a_m
\end{equation}
and the coefficients $\beta_m$ are determined by the Bogoliubov 
rotation $\mathcal{R}_{h,\gamma}$ of Eq.~\eqref{eq:bogoliubov_rot} 
that relates $\boldsymbol{\eta}$
and $\boldsymbol{\check a}$. One can check that the energy of this 
boundary mode $\check b$ tends to zero as $L\to\infty$. Therefore, in 
the thermodynamic limit, the states $\ket{\emptyset}$ and $\check 
b\ket{\emptyset}$ are degenerate and the ground state
\begin{equation}\label{eq:gs_obc}
\ket{{\rm GS}}=\frac{\ket{\emptyset}\pm \check b\ket{\emptyset}}{\sqrt{2}}
\end{equation}
spontaneously breaks the $\mathbb{Z}_2$ parity symmetry since $\langle \sigma_j^x\rangle\neq 0$.

The state~\eqref{eq:gs_obc} is a linear combination of Slater determinants that in general does not satisfy Wick theorem. This 
means that its reduced density matrix $\rho_A$ is not Gaussian and
the usual methods to univocally determine it in terms of the two-point fermionic 
correlations~\cite{peschel}, see also Appendix~\ref{app:gaussian_rules}, cannot be applied in this case. For example, let us take $h=0$ and $\gamma=1$. One of the two degenerate symmetry-broken ground states~\eqref{eq:gs_obc} is 
\begin{equation}\label{symbstate}
    \ket{{\rm GS}; h=0, \gamma=1} = \frac{1 + \check a_1}{\sqrt 2}\ket \emptyset = \ket{\rightarrow \rightarrow \rightarrow \cdots}.
\end{equation}
In this configuration, it is easy to see that a non-zero order parameter implies a non-zero odd-point fermionic correlation function,
\begin{equation}
    \langle \sigma_j^x \rangle_{h=0, \gamma=1} = \langle \check a_1 \rangle_{h=0, \gamma=1} = 1,
\end{equation}
and, therefore, Wick theorem is not satisfied. 

\subsection{Generalised Wick theorem}
For symmetry-breaking states of the form~\eqref{eq:gs_obc}, it is actually possible to obtain a generalised version of Wick theorem that
allows to express any correlator in terms of the one-point and two-point fermionic correlation functions. The rules that we must apply depend on the behaviour of the operators under the parity transformation~\eqref{eq:parity_op}.

For even operators $[\mathcal{O}_e, P] = 0$, which correspond to even-point fermionic correlators, we have
\begin{equation}
\langle \mathcal{O}_e \rangle=\frac{1}{2}\bra{\emptyset}\mathcal{O}_e\ket{\emptyset}+\frac{1}{2}\bra{\emptyset}\check b \mathcal{O}_e \check b \ket{\emptyset},
\end{equation}
where we took into account that the odd-point functions $\bra{\emptyset} \check b \mathcal{O}_e\ket{\emptyset}=\bra{\emptyset} \mathcal{O}_e \check b \ket{\emptyset}=0$ since $\ket{\emptyset}$ is a Slater determinant and we can apply Wick theorem. Using the definition~\eqref{eq:boundary_mode} of the boundary mode $\check b$, we find
\begin{equation}\label{eq:genwickeven}
\langle \mathcal{O}_e\rangle =\frac{1}{2}\bra{\emptyset}\mathcal{O}_e\ket{\emptyset}+\frac{1}{2}\sum_{m, m'}\beta_m\beta_{m'}\bra{\emptyset} \check a_m \mathcal{O}_e \check a_{m'}\ket{\emptyset}. 
\end{equation}
Observe that, in the correlators that appear in the right hand side of this equation, we can apply the usual Wick theorem and they can be expressed in terms of the two-point fermionic correlation matrix
\begin{equation}\label{eq:Gamma_0}
 (I+i\Gamma_0)_{mm'}=\bra{\emptyset}\check a_m \check a_{m'}\ket{\emptyset}.
\end{equation}
The only extra ingredient are the coefficients $\beta_m$, which determine the boundary mode $\check b$ of Eq.~\eqref{eq:boundary_mode} and actually correspond to the 
one-point correlation functions 
\begin{equation}\label{eq:one_point}
\langle \check a_m\rangle = \frac{1}{2}\bra{\emptyset}\{\check a_m, \check b\}\ket{\emptyset}=\beta_m.
\end{equation}
For odd operators, $\{ \mathcal{O}_o, P\} = 0$, 
which correspond to odd-point fermionic correlations, one can check that the presence of the boundary mode makes them non-zero,
\begin{equation}\label{eq:genwick}
    \langle \mathcal{O}_o \rangle = \frac 1 2 \bra{\emptyset}\{\mathcal{O}_o, \check b\}\ket{\emptyset} = \frac 1 2 \sum_m \beta_m \bra{\emptyset} \{\mathcal{O}_o,  \check a_m\}\ket{\emptyset},
\end{equation}
where $\{\mathcal{O}_o,  \check a_m\}$ is an even operator and its expectation value in the vacuum $\ket{\emptyset}$ can be computed from the usual Wick theorem. 

\subsection{Reduced density matrix}

\begin{figure}[t]
  \centering
  \includegraphics[width=0.75\textwidth]{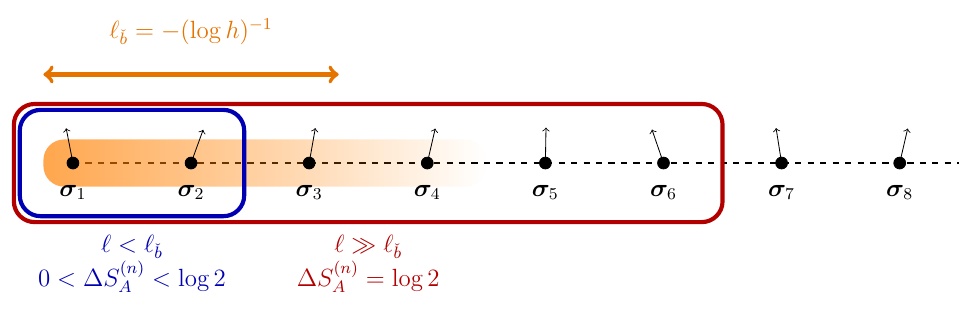}
  \caption{To study the entanglement asymmetry in a quench of the XY spin chain with OBC from the ground state~\eqref{eq:gs_obc}, we consider an interval starting from the edge where the boundary mode $\check b$ is localised. This mode has a characteristic length $\ell_{\check b}\sim -(\log h)^{-1}$. The asymmetry of~\eqref{eq:gs_obc} depends on the fraction of the mode that is contained in the interval, tending to $\log 2$ when $\ell\gg \ell_{\check b}$.}
  \label{fig:asgsobc}
\end{figure}

In the light of the previous discussion, for states of the form~\eqref{eq:gs_obc}, it is in principle possible to use the generalised Wick theorem of Eqs. \eqref{eq:genwickeven} and \eqref{eq:genwick} for even and odd correlators to reconstruct the reduced density matrix of any subsystem $A$ from the observables with support on it through the expression
\begin{equation}\label{eq:rhoa_generic}
 \rho_A=\sum_{\mathcal{O}\in A}\frac{1}{2^\ell}\langle \mathcal{O}\rangle \mathcal{O}.
\end{equation}
The problem of this formula is that, since the number of operators in a subsystem grows exponentially with its size, one cannot use it to study large subsystems. However, in Ref.~\cite{jr-22}, it is proposed an effective expression of $\rho_A$ for certain states that break parity symmetry. That ansatz can be generalised to the ground state of Eq.~\eqref{eq:gs_obc} for which we find 
\begin{equation}\label{eq:ansatz}
\rho_A=\rho_{A, e}+\frac{1}{2}\langle \check b_A\rangle \left(\check b_A \rho_{A,e} +\rho_{A,e} \check b_A\right),
\end{equation}
provided $A$ is an interval of contiguous spins starting from the edge as  in Fig.~\ref{fig:asgsobc}. In this ansatz,
$\rho_{A,e}$ is the even part of $\rho_A$ under parity $P$, which is given by the sum of two Gaussian density matrices
\begin{equation}
\rho_{A, e}=\frac{1}{2}\left(\rho_0+\tilde \rho_{0}\right),
\end{equation}
where $\rho_0$ and $\tilde \rho_{0}$ are the reduced density matrices to $A$ of the states $\ket{\emptyset}$ and $\check b\ket{\emptyset}$ respectively. We have numerically checked that, when we take the thermodynamic limit $L\to\infty$ (semi-infinite chain), $\rho_0\approx \tilde{\rho}_0$; therefore, in that case, we assume that both $\rho_0$ and $\tilde{\rho}_0$ are given by the restriction to $A$, $\Gamma_0^A$, of the correlation matrix $\Gamma_0$ in Eq.~\eqref{eq:Gamma_0}. The operator $\check b_A$ in Eq.~\eqref{eq:ansatz} is the (renormalised) restriction of the boundary mode $\check b$ of Eq.~\eqref{eq:boundary_mode} to the subsystem $A$,
\begin{equation}
\check b=\langle \check b_A\rangle \check b_A+ \langle \check b_B\rangle \check b_B.
\end{equation}
One can check that Eq.~\eqref{eq:ansatz} for $\rho_A$ correctly reproduces all the expectation values of even and odd operators predicted by the generalised Wick theorem~\eqref{eq:genwickeven} and \eqref{eq:genwick} and, importantly, it allows to efficiently compute the entanglement asymmetry of the parity, as we will see in Sec.~\ref{sec:ent_asymm_obc}. 

\subsection{One and two-point correlation functions}

\textbf{In the ground state:} In order to compute the one and two-point correlation functions of Eqs.~\eqref{eq:one_point} and \eqref{eq:Gamma_0} in the ground state~\eqref{eq:gs_obc}, the knowledge of the matrix $\mathcal{R}_{h,\gamma}$ of Eq.~\eqref{eq:bogoliubov_rot} that diagonalises the fermionic  Hamiltonian~\eqref{eq:kitaev_chain} is sufficient. In fact, the vacuum $\ket{\emptyset}$ is characterised by the absence of excitations, i.e. $\langle \eta_k^\dagger \eta_k\rangle=0$, and therefore
\begin{equation}
    \langle \boldsymbol{\eta}^\dagger \boldsymbol{\eta} \rangle = \mathcal{C}_0 \text{ with } \mathcal{C}_0 = {\rm diag}(1, 0, \dots, 1, 0).
\end{equation}
Consequently, by expressing the Majorana fermions $\check a_m$ in terms of the Bogoliubov operators $\eta_k$, $\eta_k^\dagger$ using the matrix $\mathcal{R}_{h,\gamma}$, one can compute the total system two-point correlation matrix~\eqref{eq:Gamma_0} as
\begin{equation}
    I + i\Gamma_0 = \langle \boldsymbol{\check a}^\dagger \boldsymbol{\check a} \rangle = \mathcal{R}_{h,\gamma}^{-1}  \mathcal{C}_0 \mathcal{R}_{h,\gamma} .
\end{equation}
On the other hand, since $\check b = \eta_{k_0}^\dagger + \eta_{k_0}$, the one-point fermionic functions $\beta_m$ of Eq.~\eqref{eq:one_point} are given by the sum of the two first lines of the matrix $\mathcal{R}_{h,\gamma}$ which correspond to the boundary mode according to Eq.~\eqref{eq:bogoliubov_rot} or, alternatively,
\begin{equation}\label{eq:one_point_bog_rot}
\beta_m=2\Re[(\mathcal{R}_{h,\gamma})_{1,m}].
\end{equation}

\textbf{After a quench:} Once we let the ground state~\eqref{eq:gs_obc} for $(h_0, \gamma_0)$, with $h_0<1$, unitarily evolve with the XY Hamiltonian~\eqref{eq:HXY} of another set of couplings $(h_f, \gamma_f)$, both the one and two-point correlation functions $\langle \check a_m(t)\rangle$  and $\langle \check a_m(t) \check a_{m'}(t)\rangle$ can be computed numerically using the Heisenberg picture. The Majorana operators $\check a_m$ can be evolved in time using the change of basis that relates the Bogoliubov modes that diagonalise the pre and post-quench Hamiltonians. Since the time evolution in the Heisenberg picture of the post-quech Bogoliubov operators $\tilde{\eta}_k$ and $\tilde{\eta}_k^\dagger$ is determined by
\begin{equation}
    \boldsymbol{\tilde\eta}(t) = \mathcal{U}(t) \boldsymbol{\tilde\eta}(0) \text{ with }\mathcal{U}(t) = {\rm diag}(e^{-i\tilde{\epsilon}_{k_0}t}, e^{i\tilde{\epsilon}_{k_0}t}, \dots, e^{-i\tilde{\epsilon}_{k_{L-1}}t}, e^{i\tilde{\epsilon}_{k_{L-1}}t}),
\end{equation}
we can then apply successive changes of basis to compute the dynamics of the correlation matrix~\eqref{eq:Gamma_0},
\begin{equation}
    I +i\Gamma_0(t) = \mathcal{R}_{h_f,\gamma_f}^{-1} \mathcal{U}^{-1}  (t)\mathcal{R}_{h_0, \gamma_0; h_f,\gamma_f}^{-1}   \mathcal{C}_0 \mathcal{R}_{h_0\gamma_0; h_f,\gamma_f}\mathcal{U}(t)\mathcal{R}_{h_f, \gamma_f},
\end{equation}
where $\mathcal{R}_{h_0, \gamma0; h_f, \gamma_f}$ is the matrix that expresses the modes of the pre-quench Hamiltonian in terms of the new ones, $\mathcal{R}_{h_0, \gamma_0; h_f, \gamma_f} = \mathcal{R}_{h_0, \gamma_0}\mathcal{R}_{h_f, \gamma_f}^{-1}$. 
Similarly, we can get the one-point functions $\beta_m(t)$ after the quench with the equation
\begin{equation}
    \beta_m(t) = \sum_{m'}2\Re\left[(\mathcal{R}_{h_0,\gamma_0})_{1,m'} (\mathcal{R}_{h_f, \gamma_f}^{-1} \mathcal{U}(t)\mathcal{R}_{h_f, \gamma_f})_{m',m}\right].
\end{equation}
 In this way, we have access to all the information about the dynamics of the symmetry-broken state~\eqref{eq:gs_obc} after the quench.

\subsection{Entanglement asymmetry}\label{sec:ent_asymm_obc}

With the results of the previous subsections, we can compute the entanglement asymmetry of the spin parity after a quench from the state~\eqref{eq:gs_obc} for a subsystem attached to the boundary of the chain, as in Fig.~\ref{fig:asgsobc}.

In general, according to its definition~\eqref{eq:def_ren_asym}, the calculation of the Rényi entanglement asymmetry requires to determine the moments of $\rho_A$ and its projection $\rho_{A, P}$ over the charged sectors of the parity operator $P$ in Eq.~\eqref{eq:parity_op}. In the eigenbasis of the parity in the subsystem $A$, $P_A$, the reduced density matrix $\rho_A$ takes the form
\NiceMatrixOptions{cell-space-limits=2mm, extra-margin=5pt}
\begin{equation}
\rho_A = 
    \begin{pNiceMatrix}[columns-width=0.3cm]
        \Block[draw, fill=blue!25, rounded-corners]{2-2}{\rho_{++}} & &\Block[draw, fill=red!25, rounded-corners]{2-2}{\rho_{+-}} & \\
        & & & \\
        \Block[draw, fill=red!25, rounded-corners]{2-2}{\rho_{-+}} & &\Block[draw, fill=blue!25, rounded-corners]{2-2}{\rho_{--}} & \\
        & & & \\
    \end{pNiceMatrix},
\end{equation}
where the charge sectors $\rho_{++}$ and $\rho_{--}$ are even, i.e. commute with the parity operator $P_A$, while the off-diagonal blocks $\rho_{+-}$ and $\rho_{-+}$ are odd, i.e. anticommute with it. Thus we can always write
\begin{equation}
    \rho_A = \rho_{A, e} + \rho_{A, o} \text{ and } \rho_{A, P} = \rho_{A, e}.
\end{equation}
Therefore, for the parity operator, the moments of $\rho_{A, P}$ are equal to the ones of the even part of $\rho_A$, $\Tr(\rho_{A, P}^n)=\Tr(\rho_{A, e}^n)$, which is a Gaussian operator determined by the restriction to $A$ of the correlation matrix $\Gamma_0$. On the other hand, since the trace of odd operators is zero, the moments of $\rho_A$ can be obtained from those of the even part of $\rho_A^n$, which we denote as $\rho_A^n|_e$, such as $\Tr(\rho_A^n)=\Tr(\rho_A^n|_e)$. Therefore, the definition of the Rényi entanglement asymmetry in Eq.~\eqref{eq:def_ren_asym} can be re-expressed in the form
\begin{equation}\label{eq:renyi_asymm_even_op}
\Delta S_A^{(n)}=\frac{1}{1-n}\left(\log \Tr(\rho_{A, e}^n)-\log \Tr(\rho_{A}^n|_e)\right).
\end{equation}

\begin{figure}[t]
    \centering
    \includegraphics[width=.5\textwidth]{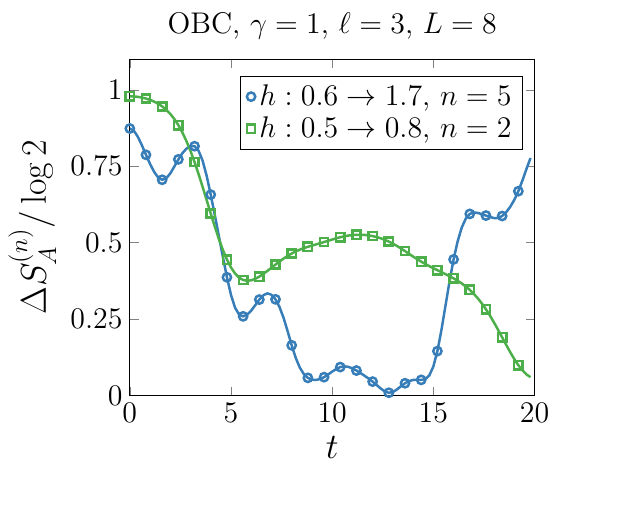}
    \caption{Numerical check of the expression~\eqref{eq:ent_asymm_obc} to compute $\Delta S_A^{(n)}$ in terms of one and two-point correlation functions after a quench in OBC from the state~\eqref{eq:gs_obc}. We consider two quenches to the ferromagnetic and paramagnetic phases and different values of the Rényi index $n$. The continuous lines correspond to Eq.~\eqref{eq:ent_asymm_obc} whereas the points have been computed using exact diagonalisation.}
    \label{fig:checkobc}
\end{figure}

In Appendix~\ref{app:ent_asymm_obc}, we employ this result together  with the ansatz of Eq.~\eqref{eq:ansatz} for $\rho_A$ to obtain an expression for the entanglement asymmetry of an interval attached to one of the boundaries in terms of the one and two-point correlation functions $\beta_m$ and $\Gamma_0$ seen before. The final formula reads
\begin{equation}\label{eq:ent_asymm_obc}
\Delta S_A^{(n)}=\frac{1}{n-1}\log\left[\frac{1}{2^n}\sum_{\{\tau_i=0, 1\}}\zeta[\{\tau_i\}]\frac{\{\Gamma_{\tau_1}^A, \dots, \Gamma_{\tau_n}^A\}}{\{\Gamma_0^A, \overset{n}{\dots}, \Gamma_0^A\}}\right],
\end{equation}
where $\Gamma_0^A$ denotes the restriction to the subsystem $A$ of 
the matrix $\Gamma_0$, $\Gamma_1^A$ is a $2\ell\times 2\ell$ matrix with entries
\begin{equation}\label{eq:gamma_1_obc}
(\Gamma_1^A)_{mm'}=(\Gamma_0^A)_{mm'}+\frac{2}{\langle \check b_A\rangle^2}\sum_{l=1}^{2\ell}\beta_l\left[(\Gamma_0^A)_{lm}\beta_{m'}-(\Gamma_0^A)_{lm'}\beta_m\right],
\end{equation}
and
\begin{equation}
    \zeta [\{\tau_i\}] = \sum_{\substack{\{w_i=0, 1\} \\ w_0=w_n=0}}\prod_{i=1}^n \Biggl(\langle\check b_A\rangle^{|w_i -w_{i-1}|} \Bigl(1 + (-1)^{\tau_i}(1-w_i-w_{i-1})\Bigl) \Biggl).
\end{equation}
The brackets $\{\Gamma_{\tau_1}, \dots, \Gamma_{\tau_n}\}$ stand for the composition of the correlations matrices of Gaussian operators, see Appendix~\ref{app:gaussian_rules} for its precise definition. While Eq.~\eqref{eq:ent_asymm_obc} becomes very cumbersome as the Rényi index $n$ increases, it simplifies to a rather simple expression for the small ones; for example, for $n=2$, we find
\begin{equation}\label{eq:ent_asymm_obc_2}
\Delta S_A^{(2)}=\log\left[1+\frac{\langle \check b_A\rangle^2}{2}+\frac{\langle \check b_A\rangle^2}{2}\frac{\{\Gamma_0^A, \Gamma_1^A\}}{\{\Gamma_0^A, \Gamma_0^A\}}\right].
\end{equation}

We can check that Eqs.~\eqref{eq:ent_asymm_obc} and \eqref{eq:ent_asymm_obc_2} are correct by comparing them with the results obtained using exact diagonalisation, as we do in Fig.~\ref{fig:checkobc}
for Rényi indexes $n=2$ and $n=5$ in two different quenches. The points are the exact numerical values of $\Delta S_A^{(n)}$ calculated using the reduced density matrix $\rho_A$ determined from the exact diagonalisation of the XY spin chain Hamiltonian~\eqref{eq:HXY} with OBC. The continuous lines correspond to Eq.~\eqref{eq:ent_asymm_obc} for $n=5$ and Eq.~\eqref{eq:ent_asymm_obc_2} in the case $n=2$.

Of course, the exact diagonalisation approach is only efficient 
for small systems, as the ones considered in Fig.~\ref{fig:checkobc}. For larger sizes,
one must resort to the expression obtained in Eq.~\eqref{eq:ent_asymm_obc} in terms of the one and two-point correlations. In what follows, we discuss the results that we find for the entanglement asymmetry using it and taking $L\to\infty$ (semi-infinite line) in quenches from the symmetry-breaking ground states~\eqref{eq:gs_obc} in the ferromagnetic region to different XY Hamiltonians both in the paramagnetic and in the ferromagnetic phases.

\begin{figure}[t]
  \centering
  \includegraphics[width=.5\textwidth]{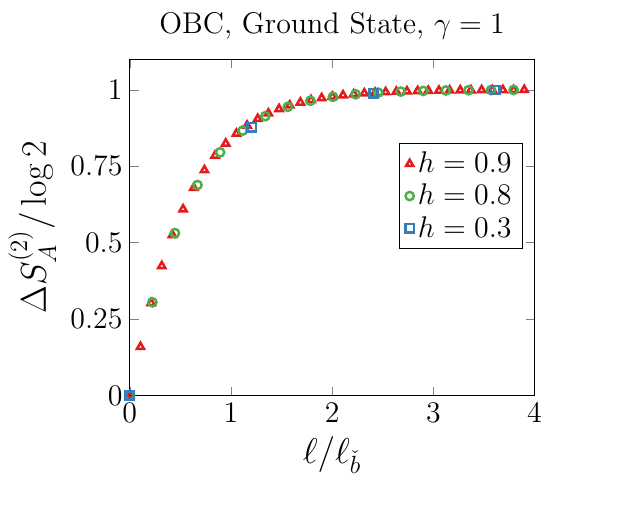}
  \caption{$n=2$ entanglement asymmetry of spin parity for the state~\eqref{eq:gs_obc} and for different values of $h<1$ and $\gamma=1$ as a function of $\ell/\ell_{\check b}$.  Here $\ell_{\check b}$ is the characteristic length of the edge mode $\check b$ localised at the boundary of the chain where the subsystem of length $\ell$ starts. The perfect overlap of the points with different $h$ shows that the entanglement asymmetry of the state~\eqref{eq:gs_obc} is a scaling function of the ratio $\ell/\ell_{\check b}$.}
\label{fig:rescaleasobc}
\end{figure}

Before proceeding, it is interesting to analyse with Eq.~\eqref{eq:ent_asymm_obc} the behaviour of the entanglement asymmetry in the initial ground state~\eqref{eq:gs_obc} as a function of the subsystem size. As Fig.~\ref{fig:asgsobc} illustrates, $A$ is an interval of contiguous spins of length $\ell$ attached to the edge of the chain where the boundary mode $\check b$ of Eq.~\eqref{eq:boundary_mode} is localised. This mode has characteristic length $\ell_{\check b}=-1/(\log h)$. Therefore, we expect that the asymmetry of the interval depends on which fraction of the boundary mode is included in the interval: if $\ell<\ell_{\check b}$, then $0< \Delta S_A^{(n)}<\log 2$ while $\Delta S_A^{(n)}\to \log 2$ when $\ell\gg \ell_{\check b}$. 
This is confirmed by Fig.~\ref{fig:rescaleasobc}, where we plot the entanglement asymmetry~\eqref{eq:ent_asymm_obc} as a function of $\ell/\ell_{\check b}$ for the ground state~\eqref{eq:gs_obc} with different values of $h$, obtaining an excellent overlap between them, signalling that the entanglement asymmetry is a scaling function of $\ell/\ell_{\check b}$.

\begin{figure}[t]
\centering
   \includegraphics[width=0.45\textwidth]{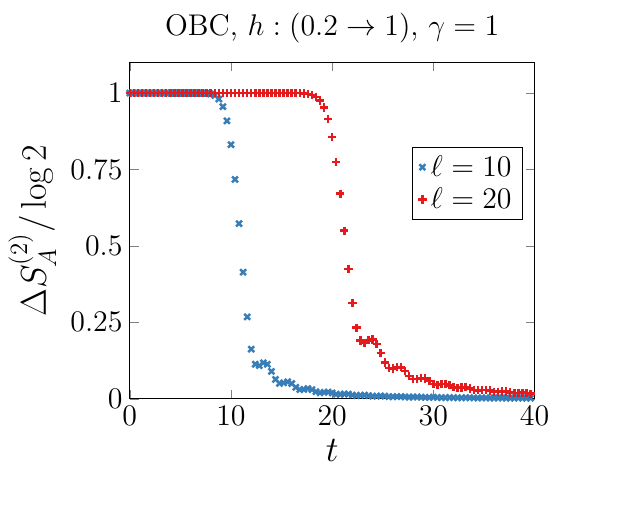} 
    \includegraphics[width=0.45\textwidth]{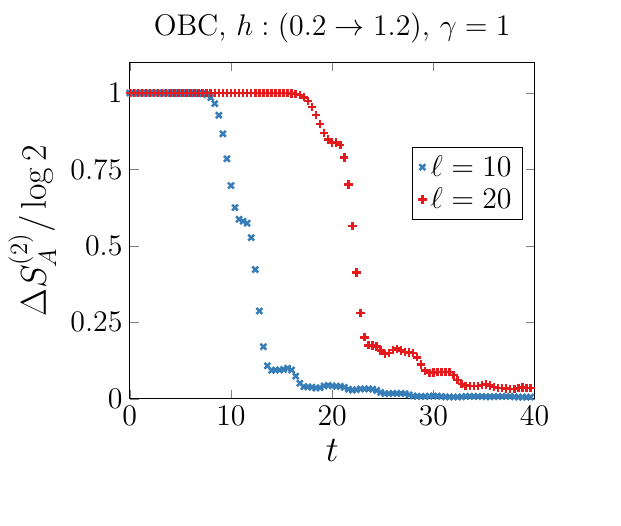} 
    \caption{Evolution of the $n=2$ entanglement asymmetry in a semi-infinite chain for a quench from the state~\eqref{eq:gs_obc} to the critical line $h=1$ (left panel) and to the paramagnetic phase $h>1$ (right panel). We take as subsystem an interval of length $\ell$ attached to the boundary. The points have been calculated using Eq.~\eqref{eq:ent_asymm_obc}. As we discuss in the main text, the asymmetry remains constant after the quench until a time $t\sim \ell/v_{\rm max}$, where $v_{\rm max}$ is the maximum velocity of the post-quench excitations, and then it drops to zero, signaling the restoration of the spin parity symmetry at large times. }%
    \label{fig:plateauobc}
\end{figure}

\textbf{Quenches to the paramagnetic phase:} When we quench to the critical point $h=1$ (left panel of Fig.~\ref{fig:plateauobc}) or to the paramagnetic phase $h>1$ (right panel of Fig.~\ref{fig:plateauobc}), we observe that the entanglement asymmetry $\Delta S_A^{(n)}$ tends to zero at large times. This implies that the $\mathbb{Z}_2$ spin parity symmetry, initially broken, is dynamically restored in the subsystem $A$ due to the \textit{melting} of the boundary mode. As evident in Fig.~\ref{fig:plateauobc}, the main feature is that the entanglement asymmetry does not instantaneously decrease after the quench, but it describes a plateau until times proportional to the subsystem length $\ell$. 

This result has a very simple interpretation: when quenching to the paramagnetic phase, the boundary mode of the pre-quench Hamiltonian can be expressed as a sum of the quasi-particle excitations of the post-quench Hamiltonian, which all have a non-zero speed of propagation. Consequently, all these modes end up escaping the subsystem. Therefore, the time at which $\Delta S_A^{(n)}$ starts to drop is roughly the time necessary for the fastest excitations to reach the end of the subsystem, that is $t\sim \ell/v_{\rm max}$ where $v_{\rm max}=\max_k|\tilde{\epsilon}_k'|$ is the maximal velocity of propagation of the post-quench Bogoliubov modes, which for $\gamma=1$ and $h\geq 1$ is one. 

\begin{figure}[h!]
  \centering
  \includegraphics[width=.5\textwidth]{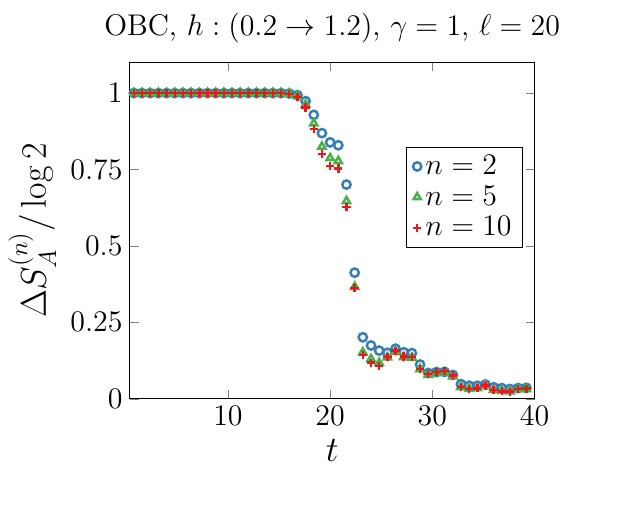}
  \caption{Time evolution in a semi-infinite chain of the Rényi entanglement asymmetry $\Delta S_A^{(n)}$ for different values of $n$ in a quench from the state~\eqref{eq:gs_obc} to the paramagnetic region. We take as subsystem an interval of fixed length $\ell=20$ starting from the boundary of the chain. The points have been obtained using Eq.~\eqref{eq:ent_asymm_obc}}
  \label{fig:reyninobc}
\end{figure}

In Fig.~\ref{fig:reyninobc}, we study the dependence of the results on the Rényi index $n$. From that plot, it is clear that it has not great relevance as the entanglement asymmetry essentially shows the same behaviour when varying its value.

\begin{figure}[t]%
    \centering
    \includegraphics[width=0.45\textwidth]{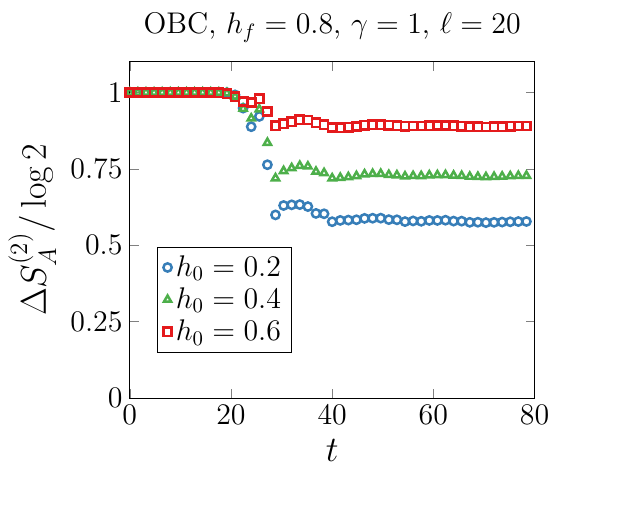}
    \includegraphics[width=0.45\textwidth]{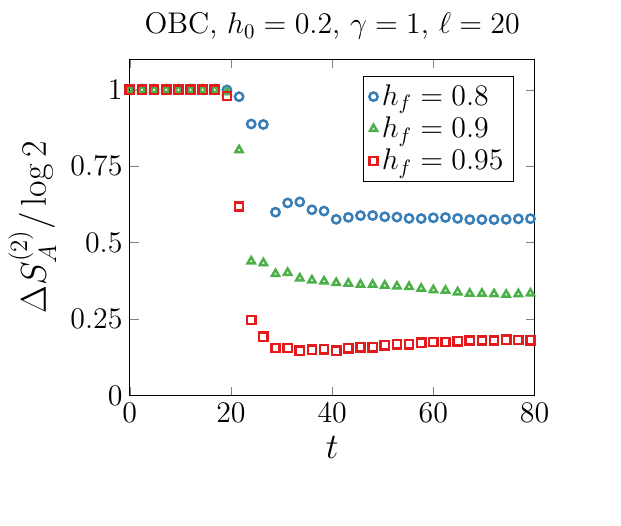}
    \caption{Dynamics of the $n=2$ entanglement asymmetry of an interval of length $\ell$ attached to the boundary of a semi-infinite chain after quenches from states of the form~\eqref{eq:gs_obc} with different values of $h_0$ to the same point in the ferromagnetic region (left panel) and from the same initial state~\eqref{eq:gs_obc} to different points in that phase (right panel). Contrary to quenches to the paramagnetic phase, cf. Fig.~\ref{fig:plateauobc}, the entanglement asymmetry does not vanishes at large times, and the spin parity symmetry is not restored, due to the existence of a boundary mode in the post-quench Hamiltonian, see also the main text.}
    \label{fig:quenchbelowobc}
\end{figure}

\textbf{Quenches to the ferromagnetic phase:} If we quench to the ferromagnetic phase, $h<1$, the spin parity symmetry is only partially restored, as shown in Fig.~\ref{fig:quenchbelowobc}. In the left panel, we observe that the farther the initial magnetic field $h_0$ is from the final one $h_f$, the smaller the stationary value of $\Delta S_A^{(n)}$ is, i.e. the more the symmetry is restored. This phenomenon is due to the existence of a boundary mode in the post-quench Hamiltonian. In fact, the pre-quench boundary mode $\check b$, defined in Eq.~\eqref{eq:boundary_mode}, can be expressed in terms of the post-quench Bogoliubov excitations as
\begin{equation}
\check b = \alpha_0 \check b_f+\sum_{j\neq 0}(\alpha_j\eta_{k_j}+\alpha_j^*\eta_{k_j}^\dagger).
\end{equation}
Therefore, it has a non-zero overlap $\alpha_0$ with the boundary mode $\check b_f$ of the post-quench Hamiltonian, which has zero velocity. Consequently, part of the initial boundary mode, which is responsible for the parity symmetry breaking at $t=0$, stays in the subsystem and the symmetry is only partially restored. In the right panel of Fig.~\ref{fig:quenchbelowobc}, we consider different quenches from the same initial state. We see that the symmetry is more restored in $A$ as we quench closer to the critical point, where it is fully restored. This can be explained by the fact that the length of the post-quench boundary mode diverges as we quench closer to $h_f=1$ and, moreover, the overlap $\alpha_0$ diminishes too.

\section{Entanglement asymmetry in a periodic chain}\label{sec:pbc}

In this section, we study the time evolution after a quench of the 
entanglement asymmetry of spin parity in a XY spin chain~\eqref{eq:HXY} with 
Periodic Boundary Conditions (PBC). In this case, the presence of the 
boundary term $H_b$, given by Eq.~\eqref{eq:boundary}, in the 
fermionic Hamiltonian~\eqref{eq:kitaev_chain} makes it non-quadratic. 
However, using the fact that the Hamiltonian~\eqref{eq:HXY} is 
invariant under parity $P$ transformations~\eqref{eq:parity_op}, we can solve this issue by 
separating the Hilbert space into two sectors, called Neveu-Schwarz 
(NS) for positive parity and Ramond (R) for the negative one. Then the 
XY Hamiltonian~\eqref{eq:HXY} can be written as  
\begin{equation}
    H= \frac{1+P}{2}H_{NS}\frac{1+P}{2} + \frac{1-P}{2}H_{R}\frac{1-P}{2},
\end{equation}
where $H_{NS}$ and $H_{R}$ are the fermionic Hamiltonian~\eqref{eq:kitaev_chain} with antiperiodic and periodic boundary conditions respectively according to Eq.~\eqref{eq:boundary}. Note that both commute with the parity $P$. Therefore, if we restrict to one of these sectors, the Hamiltonian is quadratic and, by performing a Bogoliubov rotation~\eqref{eq:bogoliubov_rot} as in the case of open boundary conditions, we can diagonalise them. For the NS sector, we find
\begin{equation}
H_{NS}=\sum_{j}\epsilon_{k_j}\left(\eta_{k_j}^\dagger \eta_{k_j}-\frac{1}{2}\right),
\end{equation}
where $\epsilon_k$ is the one-particle dispersion relation introduced in Eq.~\eqref{eq:disp_rel}. In this case, the momenta $k_j$ satisfy the quantization condition
\begin{equation}
k_j=\frac{2\pi(j-\frac{1}{2})}{L},\quad j=0, \dots, L-1.
\end{equation}
For the R sector, we have
\begin{equation}
H_R=\sum_{j\neq 0}\epsilon_{p_j}\left(\eta_{p_j}^\dagger\eta_{p_j}-\frac{1}{2}\right)-2(1-h)\left(\eta_{0}^\dagger\eta_0-\frac{1}{2}\right)
\end{equation}
with the momenta $p_j$ satisfying
\begin{equation}
p_j=\frac{2\pi j}{L},\quad j=0, \dots, L-1.
\end{equation}

These two sectors define two different vacuums, $\ket{\emptyset}_{NS}$ and $\ket{\emptyset}_R$, such that
\begin{equation}
    \eta_{k_j} \ket{\emptyset}_{NS} = 0 \text{ and } \eta_{p_j} \ket{\emptyset}_{R} = 0, \forall j.
\end{equation}
If the chain has a finite size, then its ground state is the vacuum 
$\ket{\emptyset}_{NS}$. In the thermodynamic limit $L\to\infty$, as 
we already mentioned, the states $\ket{\emptyset}_{NS}$ and $\eta_0^\dagger \ket{\emptyset}_R$ become degenerate in the ferromagnetic phase $h<1$ and the ground state is the linear superposition
\begin{equation}\label{eq:pbcstate}
    \ket \Psi = \frac{\ket{\emptyset}_{NS} + \eta_0^\dagger \ket{\emptyset}_R}{\sqrt 2},
\end{equation}
which spontaneously breaks the $\mathbb{Z}_2$ parity symmetry \eqref{parityfermions}. As happens for the ground state~\eqref{eq:gs_obc} with OBC, the state~\eqref{eq:pbcstate} does not satisfy the Wick theorem.  However, the main difference is that the odd part in Eq.~\eqref{eq:pbcstate} consists of an excitation over a vacuum (R vacuum) different from the the vacuum of the even part (NS vacuum). 
Thus one cannot use the ansatz~\eqref{eq:ansatz} for $\rho_A$ in OBC. 
We have to take an alternative route based on cluster decomposition to calculate the entanglement asymmetry in a quench from the state~\eqref{eq:pbcstate}. 

In the thermodynamic limit, the correlation functions of even operators $\mathcal{O}_e$ are equal for $\ket{\emptyset}_{NS}$ and $\eta_0^\dagger\ket{\emptyset}_R$~\cite{cef-12-1}. Therefore, the even-point fermionic correlations functions of the state~\eqref{eq:pbcstate} can be calculated using Wick theorem and, consequently, are fully characterised by the two-point correlation matrix $\Gamma_0$ defined in Eq.~\eqref{eq:Gamma_0}, which can be determined using the same procedure as in OBC. In the thermodynamic limit, it takes the following form after the quench~\cite{cef-12-1}
\begin{equation}\label{eq:cor1pbc}
(\Gamma_0(t))_{jj'}=\int_{-\pi}^{\pi}\frac{dk}{2\pi}
\mathcal{G}(k, t) e^{ik(j-j')}, \quad j, j'=1, 2,\dots,
\end{equation}
where $\mathcal{G}(k, t)$ is the $2\times 2$ matrix
\begin{equation}
\mathcal{G}(k, t)=\left(\begin{array}{cc}
 f(k) & -g(k)\\
  g(-k) & -f(k)
\end{array}\right)
\end{equation}
with
\begin{equation}
f(k)=i\sin\Delta_k\sin(2\tilde{\epsilon}_kt),\quad 
g(k)=e^{-i\theta_k^f}(\cos\Delta_k+i\sin\Delta_k\cos(2\tilde{\epsilon}_kt)).
\end{equation}
Here $\Delta_k$ is the difference $\Delta_k = \theta_k^f- \theta_k^0$ between the Bogoliubov angles of the post and pre-quench Hamiltonians with couplings $(h_f, \gamma_f)$ and $(h_0, \gamma_0)$ respectively and $\tilde{\epsilon}_k$ is the dispersion relation of the post-quench Hamiltonian. The Bogoliubov angle for arbitrary couplings $h$, $\gamma$ is given by 
\begin{equation}
e^{i\theta_k}=\frac{\cos k - h +i\gamma\sin k}{\sqrt{(h-\cos k)^2+\gamma^2\sin^2 k}}.
\end{equation}

On the other hand, the calculation of odd-point fermionic correlations
in the state~\eqref{eq:pbcstate} is a rather difficult task. However, following the ideas in Ref.~\cite{fc-10}, we can use the cluster decomposition principle to determine them from the even-point ones, exploiting the fact that the latter are equal in both parity sectors in the thermodynamic limit and are given by the matrix $\Gamma_0$. In fact, let us take an arbitrary odd operator $\mathcal{O}_o$ with support in the subsystem $A$. We further consider the auxiliary spin operator $\sigma_r^x$ on the site $r$ that does not belong to the interval $A$. If we take the even correlation function $\langle \mathcal{O}_o\sigma_r^x\rangle$, then we expect than in the infinite distance separation, $r\to\infty$, the spin at the site $r$ is uncorrelated from the spins in the subsystem $A$
and
\begin{equation}
    \langle \mathcal{O}_o \sigma^x_r \rangle \underset{r\to \infty}{=}\langle \mathcal{O}_o \rangle \langle \sigma_r^x \rangle.
 \end{equation}
 Therefore,
 \begin{equation}
    \langle \mathcal{O}_o \rangle = \pm \lim_{r\to \infty} \frac{\langle \mathcal{O}_o \sigma^x_r \rangle}{\sqrt{\langle \sigma^x_1 \sigma^x_r \rangle}}.
\end{equation}
Using this idea, in Appendix \ref{appendix:cluster_decomposition}, we show that it is possible to obtain the odd part of $\rho_A$ for the state~\eqref{eq:pbcstate}. Applying some of the methods to compute the entanglement entropy of disjoint intervals in the XY spin chain \cite{fc-10}, we find that $\rho_A$ can be expressed in terms of two Gaussian operators $\rho_0$ and $\rho_1$
\begin{equation}\label{eq:rdmpbc}
    \rho_{A} = \rho_0 + \langle \sigma_1^x \rangle \sigma_1^x \rho_1,
\end{equation}
which are respectively determined by the restriction to $A$, $\Gamma_0^A$, of the correlation matrix \eqref{eq:cor1pbc} and by $\Gamma_1^A$, defined as 
\begin{equation}\label{eq:cor2pbc}
    (I + i\Gamma_1^A)_{mm'} = \frac{\langle \check a_m \check a_{m'} \sigma^x_1\rangle}{\langle \sigma^x_1 \rangle}, \quad m,m'\in A.
\end{equation}
As we show in Appendix~\ref{appendix:cluster_decomposition}, the latter can be obtained from $\Gamma_0$ using cluster decomposition as a Schur complement.

\subsection{Entanglement asymmetry}

From the expression \eqref{eq:rdmpbc} of the reduced density matrix $\rho_A$ for the ground state~\eqref{eq:pbcstate}, we can derive using Eq.~\eqref{eq:renyi_asymm_even_op} an efficient formula to calculate the entanglement asymmetry of the spin parity symmetry for a large subsystem. Its obtention involves some tedious algebraic manipulations, which are described in detail in Sec.~\ref{app:ent_asymm_pbc}. The final result reads
\begin{equation}\label{eq:ent_asymm_pbc}
    \Delta S_A^{(n)} = \frac{1}{n-1}\log \left[\sum_{\{\chi_i=0,1\}}\frac{1+(-1)^{s_n}}{2}\langle \sigma^x \rangle^{s_n}  \frac{\{P_1^{s_1}\Gamma_{\chi_1}^AP_1^{s_1}, \cdots, P_{1}^{s_n}\Gamma_{\chi_i}^AP_{1}^{s_n}\}}{\{\Gamma_0^A, \overset{n}{\dots}, \Gamma_0^A\}}\right],
\end{equation}
where $\Gamma_0^A$ and $\Gamma_1^A$ are the restriction to $A$ of the correlation matrices \eqref{eq:cor1pbc} and \eqref{eq:cor2pbc} respectively, $s_n = \sum_{j=1}^n\chi_j$ and $P_{1}$ is the diagonal matrix $P_1 = {\rm diag}(-1, 1, \cdots, 1)$. The Rényi-2 asymmetry has a simpler expression
\begin{equation}
\Delta S_A^{(2)}=\log\left[1+\frac{\{\Gamma_1^A, P_1\Gamma_1^AP_1\}}{\{\Gamma_0^A, \Gamma_0^A\}}\right].
\end{equation}

\begin{figure}[t]
    \centering
    \includegraphics[width=.5\textwidth]{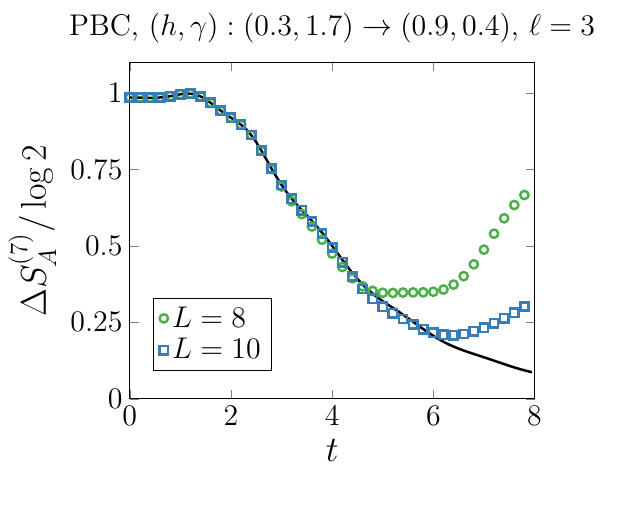}
    \caption{Numerical check of the formula~\eqref{eq:ent_asymm_pbc} to calculate $\Delta S_A^{(n)}$ after a quench in PBC from the state~\eqref{eq:pbcstate} in terms of the one and two-point fermionic correlation functions. The points are the value of $\Delta S_A^{(n)}$ obtained using exact diagonalisation for two chains of different length $L$. The continuous line corresponds to Eq.~\eqref{eq:ent_asymm_pbc}, which is valid in the thermodynamic limit. This explains the discrepancies with the points at times in which the finite-size effects become relevant.}
    \label{fig:checkpbc}
\end{figure}

As we did for OBC, we check that the formula of Eq.~\eqref{eq:ent_asymm_pbc} produces the correct entanglement asymmetry by comparing with calculations done using exact diagonalisation for small systems. In Fig.~\ref{fig:checkpbc}, we analyse the time evolution of $\Delta S_A^{(n)}$ for $n=7$ in a particular quench from the state~\eqref{eq:pbcstate}. The continuous line corresponds to Eq.~\eqref{eq:ent_asymm_pbc} while the points have been obtained with exact diagonalisation for two different total sizes $L$ of the chain. Observe that, while  the curve and the points match just after the quench, they deviate at large times. The reason is that Eq.~\eqref{eq:ent_asymm_pbc} gives $\Delta S_A^{(n)}$ in the thermodynamic limit $L\to\infty$. In fact, recall that this expression relies on the cluster decomposition between the spins of subsystem $A$ and an auxiliary spin. Therefore, it is expected that our method only agrees with the exact diagonalisation results at early times and small correlations lengths, before finite-size effects, as the revivals of Fig.~\ref{fig:checkpbc}, appear.

The rest of this section is devoted to discussing the results that we obtain for the entanglement asymmetry after quenches from the symmetry-breaking ground state~\eqref{eq:pbcstate} in the ferromagnetic phase to different points of the $(h,\gamma)$-plane of the XY spin chain. We first consider the case of a subsystem made of a single site in order to compare the entanglement asymmetry with the usual order parameter $\langle \sigma_1^x\rangle$. We will then take larger subsystem sizes and obtain an analytic expression that describes the time evolution of the asymmetry at leading order.  

\paragraph{One-site subsystems --- comparison with the order parameter:}
As we already pointed out in Sec.~\ref{sec:xy_chain}, in the thermodynamic limit, the breaking of the $\mathbb{Z}_2$ symmetry associated to spin parity can be detected with the standard order parameter $\langle \sigma_1^x\rangle$. It is then interesting to compare the results for $\langle \sigma_1^x\rangle$ with those of the asymmetry for a subsystem of one site. By cluster decomposition, we can indeed get $\langle \sigma_1^x \rangle^2$ by computing the expectation value of a string of Majorana fermions as described in Ref.~\cite{cef-12-1}.

According to Eq.~\eqref{eq:rhoa_generic}, the reduced density matrix of a one site subsystem is simply
\begin{equation}
    \rho_A = \frac 1 2 \left[ I + \langle \sigma_1^z \rangle \sigma_1^z + \langle \sigma_1^x \rangle \sigma_1^x + \langle \sigma_1^y \rangle \sigma_1^y\right]
\end{equation}
and its projection on the $\mathbb{Z}_2$ parity sectors is
\begin{equation}
    \rho_{A, P} = \frac 1 2 \left[ I + \langle \sigma_1^z \rangle \sigma_1^z\right].
\end{equation}
Using now the anticommutation relations of Pauli matrices, one can derive
\begin{equation}\label{eq:ent_asymm_one_site}
    \Delta S^{(n)}_A = H^{(n)}\left(\langle \sigma_1^z \rangle\right) - H^{(n)}\left(\sqrt{\langle \sigma_1^x \rangle^2+ \langle \sigma_1^y \rangle^2+ \langle \sigma_1^z \rangle^2}\right),
\end{equation}
where the function $H^{(n)}(x)$ is defined in Eq.~\eqref{eq:hncircle}. In this expression, we can  take the limit $n\to1$ according to Eq.~\eqref{eq:hnto1}. Observe that the asymmetry of one site is not exactly equivalent to the order parameter $\langle \sigma_1^x\rangle$ since it takes into account the symmetry breaking in both directions $x$ and $y$. In fact, $\Delta S_A^{(n)}=0$ if and only if both $\langle \sigma_1^x \rangle$ and $\langle \sigma_1^y \rangle$ are zero.
\begin{figure}[t]%
    \centering
    \includegraphics[width=7cm]{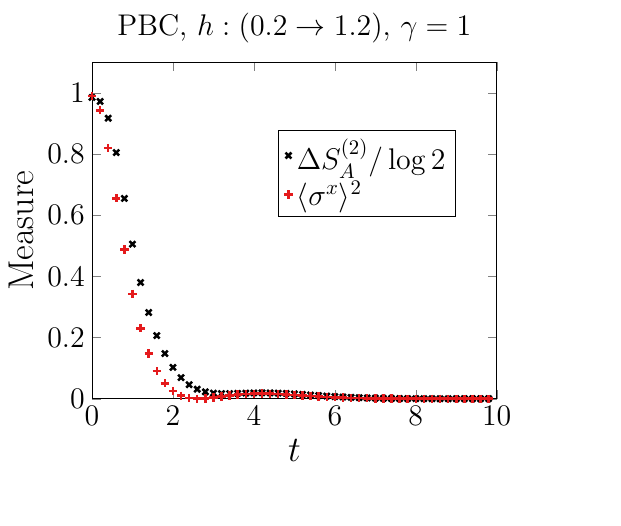} 
    \includegraphics[width=7cm]{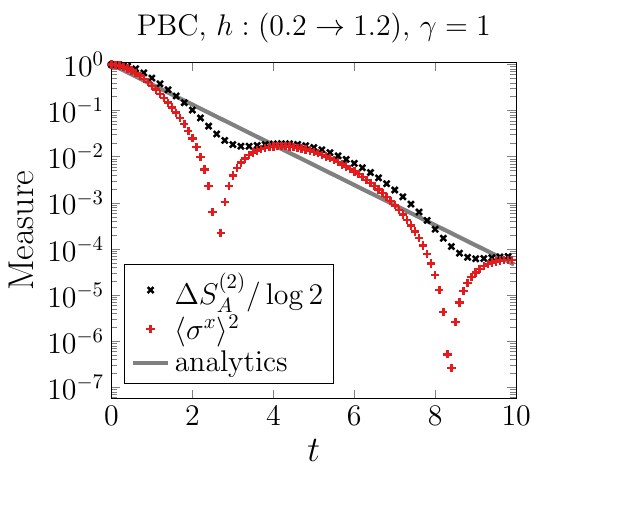} 
    \caption{Left panel: comparison between the order parameter $\langle \sigma_1^x\rangle^2$ and the entanglement asymmetry $\Delta S_A^{(2)}$ of spin parity for a single site after a quench from a particular symmetry-breaking ground state~\eqref{eq:pbcstate} to the paramagnetic region. Right panel: same plot taking logarithmic scale in the $y$ axis. The entanglement asymmetry $\Delta S_A^{(2)}$ has been obtained using Eq.~\eqref{eq:ent_asymm_one_site} in terms of the one-site magnetisations $\langle \sigma^{x}_1\rangle$, $\langle \sigma^{y}_1\rangle$ and $\langle \sigma^{z}_1\rangle$  which can be determined numerically using the methods of Ref.~\cite{cef-12-1}. The solid line is the asymptotic behaviour~\eqref{eq:sxanalytics} of $\langle \sigma_1^x\rangle^2$ for large times.}
    \label{fig:as1sx}%
\end{figure}
In the left panel of Fig.~\ref{fig:as1sx}, we compare both quantities for a particular quench. If we take logarithmic scale in the $y$ axis, as we do in the right panel Fig.~\ref{fig:as1sx}, their differences are more obvious. We can see that, while the order parameter $\langle \sigma_1^x\rangle^2$ periodically vanishes at certain times due to the oscillations between the $x$ and $y$ components of the spin, the asymmetry does not since it takes both directions of symmetry breaking into account.
This is a physical important result that shows the power of the asymmetry compared to traditional observables: {\it the symmetry is never restored at intermediate time although the order parameter vanishes}. 

Despite the differences between the order parameter $\langle \sigma_1^x\rangle^2$ and the asymmetry, a remarkable result is that the leading term at late times is given by the same exponential decay for both quantities. In Refs.~\cite{cef-11, cef-12-1}, it was found that 
\begin{equation}\label{eq:sxanalytics}
    \langle \sigma_1^x (t) \rangle^2 \underset{t\to \infty} {\sim}\exp(2t\int_{0}^\pi \frac{dk}{\pi} |\tilde{\epsilon}_k'|  \log |\cos \Delta_k |),
\end{equation}
which corresponds to the solid line in the right panel of Fig.~\ref{fig:as1sx}.

\begin{figure}[t]
    \centering
    \includegraphics[width=0.42\textwidth]{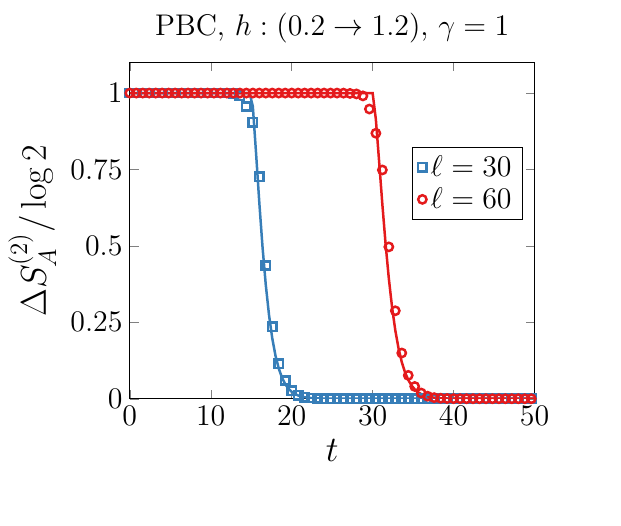}
    \includegraphics[width=0.42\textwidth]{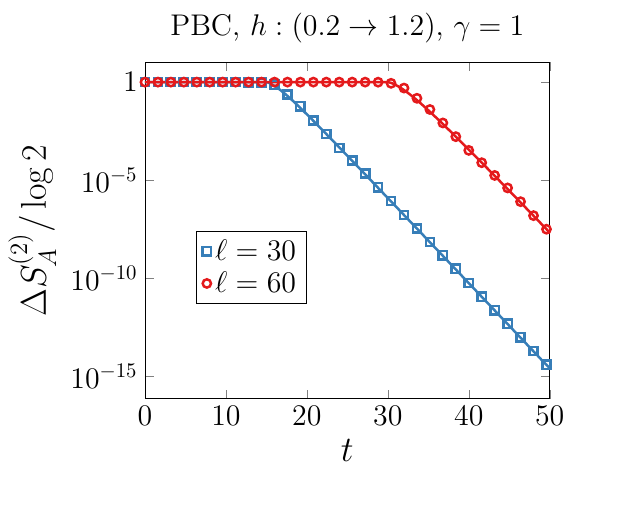}
    \includegraphics[width=0.42\textwidth]{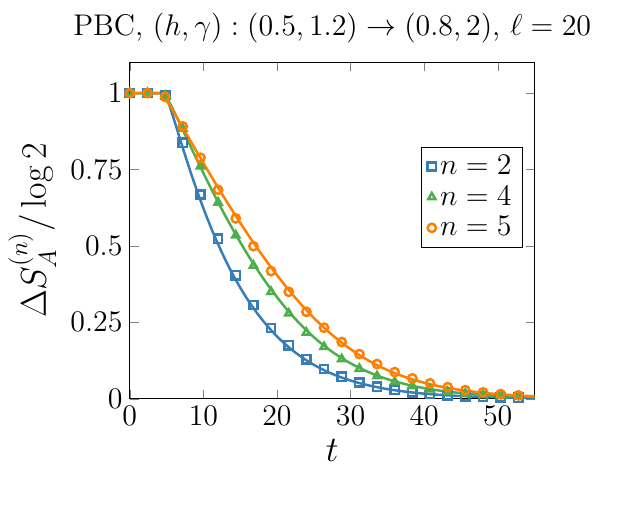}
    \includegraphics[width=0.42\textwidth]{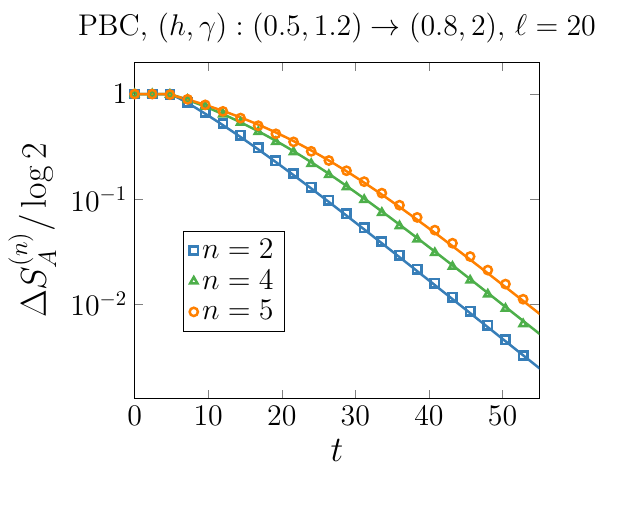}
    \caption{Left panels: Time evolution in an infinite chain of the Rényi entanglement asymmetry of spin parity for different quenches in which the chain is initially prepared in the symmetry breaking ground state~\eqref{eq:pbcstate} for a specific set of couplings $(h_0, \gamma_0)$ in the ferromagnetic region and then is let evolved with the XY Hamiltonian~\eqref{eq:HXY} with different couplings $(h_f, \gamma_f)$ in the paramagnetic (upper panel) and in the ferromagnetic (lower panel) phases. Each panel corresponds to a particular quench, in the upper one we take two subsystem lengths $\ell$ and fixed Rényi index $n$ and vice versa in the lower one. Right panels: same plots taking logarithmic scale in the $y$ axis. The points have been obtained using Eq.~\eqref{eq:ent_asymm_pbc} while the solid curves correspond to the analytic conjecture \eqref{eq:ent_asymm_analytic} in which the free parameter $C_{h_0, \gamma_0;h_f,\gamma_f}$ has been adjusted to obtain the best agreement with the numerical points.}%
    \label{fig:pbc_quenches}%
\end{figure}

\paragraph{Larger subsystems:} 
In Fig.~\ref{fig:pbc_quenches}, we study the time evolution of the entanglement asymmetry for larger subsystems in different quenches starting from an initial state of the form~\eqref{eq:pbcstate} that spontaneously breaks the $\mathbb{Z}_2$ spin parity symmetry and letting it evolve with an XY spin chain Hamiltonian with different couplings $h$ and $\gamma$ both in the paramagnetic and in the ferromagnetic phases. In all the cases, we observe a plateau in the asymmetry at early times after the quench followed by an exponential decay. In fact, in the upper left panel of Fig.~\ref{fig:pbc_quenches}, we study a particular quench to the paramagnetic phase for two different subsystem lengths; in the upper right panel, we consider the same quench but taking logarithmic scale in the $y$ axis to confirm the exponential relaxation to zero of the entanglement asymmetry. In the case of OBC, we explain the existence of the plateau in terms of the boundary mode present in the subsystem $A$. In PBC, even if we do not have the boundary mode picture, by analogy, we can think that the quasi-particle excitations generated after the quench emerge from the two end-points of the subsystem. This would explain that the length of the plateau is twice shorter, $\sim \ell/(2v_{\rm max})$, than in OBC (recall that, in that case, the interval is attached to the boundary and the quasi-particles can only emerge from one point). 

We find that the exponential decay is rather universal for all the 
quenches. The change of the initial and final couplings $(h_0, 
\gamma_0)$ and $(h_f,\gamma_f)$ only affects the decay rate, which is 
independent of the subsystem size $\ell$ and the Rényi index $n$. This observation is crucial since, as we have seen above, both the asymmetry of a single site and 
 the order parameter $\langle \sigma_1^x\rangle^2$ show the same exponential decay at late times, which is described by Eq.~\eqref{eq:sxanalytics}. We can use this fact to conjecture an effective analytic expression for the asymmetry of 
 a subsystem of length $\ell$ after the quench. We find that it is actually well
 reproduced by the formula
\begin{equation}\label{eq:ent_asymm_analytic}
\Delta S_A^{(n)}(t)\approx \frac{\log 2}{\log C_n}\log\left[1+(C_n-1){\rm exp}\left(\int_{0}^\pi \frac{dk}{\pi}\max(2|\tilde{\epsilon}_k'|t-\ell +\mathcal{O}(1), 0)\log|\cos\Delta_k|\right)\right],
\end{equation}
where $C_n = 2+(n-2)C_{h_0, \gamma_0; h_f, \gamma_f}$ and $C_{h_0, \gamma_0; h_f, \gamma_f}$ is an unknown coefficient that depends on the couplings of the pre and post-quench Hamiltonian. Observe that Eq.~\eqref{eq:ent_asymm_analytic} satisfies the required properties, for $t<l/2v_{max}$ it simplifies to $\Delta S_A^{(n)}(0)=\log 2$ while at larges times it exponentially decays at leading order as the order parameter~\eqref{eq:sxanalytics},
\begin{equation}\label{eq:asymmetry_late}
\Delta S_A^{(n)}(t)\underset{t\to \infty} {\sim}{\rm exp}\left(2t\int_{0}^\pi \frac{dk}{\pi}|\epsilon_k'|\log|\cos\Delta_k|\right),
\end{equation}
which is independent of the subsystem lenght $\ell$ and Rényi index $n$, and in particular is valid  in the limit $n\to 1$.
The solid curves in all the plots of Fig.~\ref{fig:pbc_quenches} correspond to the analytic expression~\eqref{eq:ent_asymm_analytic}. For each quench, the value of the unknown coefficient $C_{h_0, \gamma_0;h_f,\gamma_f}$ has been adjusted by hand to obtain the best
agreement with the numerical points calculated using Eq.~\eqref{eq:ent_asymm_pbc}.

\begin{figure}[t]
  \centering
  \includegraphics[width=.325\textwidth]{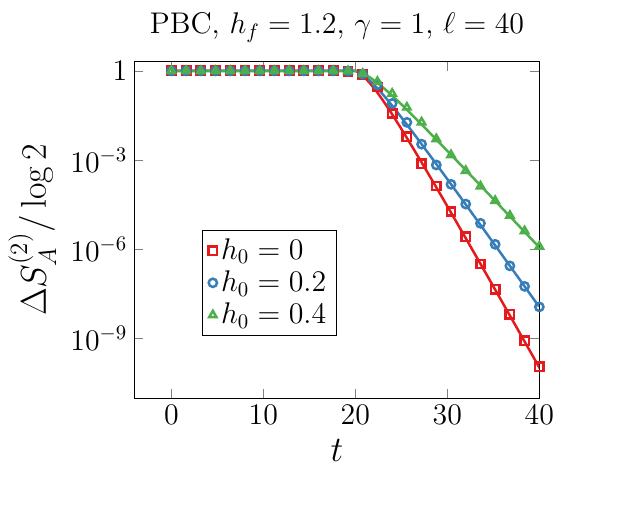}
  \includegraphics[width=0.325\textwidth]{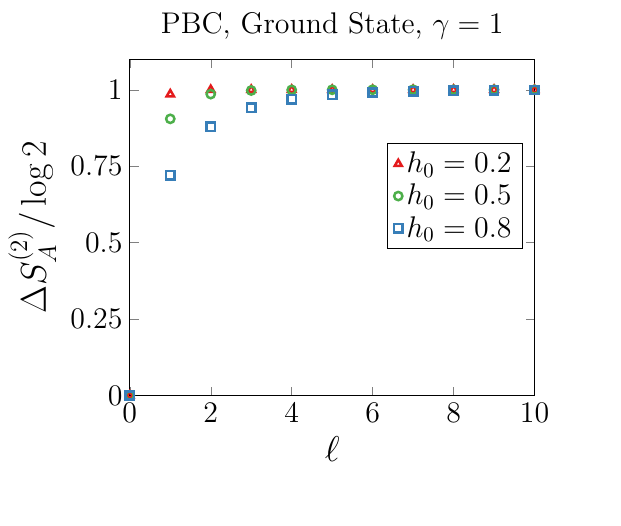} 
    \includegraphics[width=0.325\textwidth]{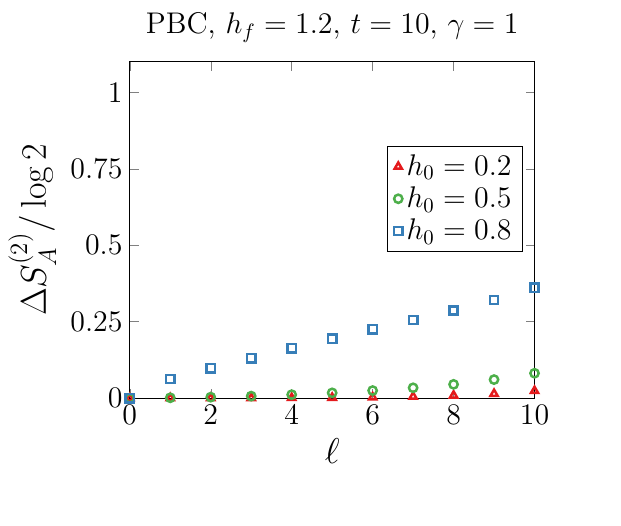} 
  \caption{Time evolution of the $n=2$ entanglement asymmetry of spin parity in a quench from ground states~\eqref{eq:pbcstate}. Left:  different $h_0$ and $\gamma_0=1$ taking the same post-quench  Hamiltonian with $h_f=1.2$ and $\gamma_f=1$, i.e., in the paramagnetic region. The points have been obtained using Eq.~\eqref{eq:ent_asymm_pbc} and the solid curves correspond to the analytic conjecture of Eq.~\eqref{eq:ent_asymm_analytic}. As can be seen, the farther the initial state is from the paramagnetic region the faster the symmetry is restored.
  Center: Initial state asymmetry for different $h_0$ and $ \gamma=1$. The points have been obtained with Eq.~\eqref{eq:ent_asymm_pbc}. 
  Right: same plot as in center but at time $t=10$ after a quench  to the Hamiltonian $(h_f, \gamma_f)=(1.2, 1)$. Comparing center and right panels, we observe that the points are inverted: at any length scale $\ell$, the symmetry is more restored at $t=10$ by the states that break it more at $t=0$, a reminiscence of the quantum Mpemba effect observed in  $U(1)$ symmetries. }
  \label{fig:pseudompemba1}
\end{figure}

\textbf{Quantum Mpemba effect?}
In the case of the $U(1)$ symmetry associated to the transverse magnetisation, a quantum version of Mpemba effect has been observed, in which the symmetry is restored faster after a quench for the initial states that break it more~\cite{amc-22}. For the $\mathbb{Z}_2$ spin parity symmetry, we cannot in principle observe the same effect since the asymmetry at time $t=0$ converges to $\log 2$ very fast as the length $\ell$ of the subsystem increases. Therefore, for large enough intervals, the spin parity symmetry is maximally broken for any initial state of the form~\eqref{eq:pbcstate} that we can consider. 
However, when studying a series of quenches to the same point in the  paramagnetic region from different initial states such as the ones plotted in the left panel of Fig.~\ref{fig:pseudompemba1}, it is very tempting to conclude that the symmetry is faster restored for the ground state corresponding to $h_0=0$ than for $h_0=0.2$ or $h_0=0.4$, which are parameters closer to the paramagnetic region. 

On the other hand, it is important to remember that a value $\log 2$ for the asymmetry of a $\ell$ site subsystem means that the symmetry is completely broken at that length scale. 
If we analyse the entanglement asymmetry of the initial states~\eqref{eq:pbcstate} as a function of the subsystem size, as we do in the middle panel of Fig.~\ref{fig:pseudompemba1}, we can see that, for any subsystem length, the entanglement asymmetry for $h_0=0.8$ is smaller than for $h_0=0.5$ and $h_0=0.2$. In the right panel of Fig.~\ref{fig:pseudompemba1}, we plot the value of entanglement asymmetry at time $t=10$ after performing a quench to the same point $(h_f, \gamma_f)=(1.2, 1)$ from the initial states considered in the middle panel.  As we see, at $t=10$, the points are inverted with respect to the ones at initial time, now the symmetry is more broken for the state corresponding $h_0=0.8$ than for $h_0=0.5$ and $h_0=0.2$. This is a phenomenon somehow reminiscent of the quantum Mpemba effect found in the $U(1)$ case in the sense that the more the spin parity symmetry is broken at any length scale, the faster it restores at any length scale. 


\section{Conclusions}\label{sec:conclusions}

In this manuscript, we have extended to finite discrete cyclic groups $\mathbb{Z}_N$ the study of the recently introduced entanglement asymmetry. Contrary to the $U(1)$ symmetry considered in Ref.~\cite{amc-22}, the entanglement asymmetry is bound by the dimension of the group. As an application, we have examined the entanglement asymmetry 
of the $\mathbb{Z}_2$ group generated by the spin parity in the XY spin chain. In particular, we have analysed its time evolution after a quench from the ferromagnetic phase, where this symmetry is spontaneously broken by the ground state in the thermodynamic limit. This extends the analysis carried out in Ref.~\cite{cef-11, cef-12-1} using the standard order parameter. While the order parameter could be employed as a measure of symmetry breaking for one-site subsystems, the longer range correlations that break the symmetry in extended subsystems are only taken into account in the asymmetry, which is then a proper quantifier of symmetry breaking for larger subsystems.
Furthermore we showed that the asymmetry is more suited to detect symmetry breaking since we found examples in which the order parameter vanishes when the symmetry is still broken. 

We were faced with the technical issue that the states that break this $\mathbb{Z}_2$ symmetry do not satisfy the Wick theorem, and the powerful techniques for Gaussian reduced density matrices employed in the case of the $U(1)$ symmetry in Refs.~\cite{amc-22} and \cite{amvc-23} cannot be directly applied here. 
We have solved this problem obtaining the reduced density matrix by using a generalised version of Wick theorem that takes into account the one-point functions in open boundary conditions and by applying the cluster decomposition principle when the chain has periodic boundary conditions. 

According to our results, the entanglement asymmetry rapidly saturates for large subsystems to its maximal value, $\log 2$, in all the ferromagnetic phase, being zero in the paramagnetic region; therefore, it acts as a topological indicator of symmetry breaking for large subsystems. The entanglement asymmetry also allows to determine when and how the $\mathbb{Z}_2$ parity symmetry is restored in a subsystem after a quench to another set of couplings of the XY spin chain. In particular, we have found that, just after the quench, the entanglement asymmetry describes a plateau of length proportional to the subsystem size and then it suddenly decreases its value. 
For open boundary conditions, if the quench is to the paramagnetic region, the asymmetry of an interval starting from the boundary drops to zero at large times and the symmetry is fully restored, while in a quench to the ferromagnetic phase, it tends to a non-zero value and the symmetry is only partially restored due to the 
boundary mode of the post-quench Hamiltonian. 
For periodic boundary conditions, the symmetry is fully restored in all the cases. Remarkably, we have found that, at large times and any Rényi index, the entanglement asymmetry shows the same exponential decay to zero as the order parameter, for which an analytic expression was derived in Refs.~\cite{cef-11, cef-12-1}. From it, we have conjectured an analytic expression for the entanglement asymmetry that effectively reproduces 
its time evolution after the quench. 

\section*{Acknowledgements}
 We thank Sara Murciano for useful discussions. FF acknowledges the financial support from the Erasmus+ program and the excellence IDEX grant of Université Paris-Saclay.  PC and FA acknowledge support from ERC under Consolidator Grant number 771536 (NEMO).

\appendix

\section*{Appendices}

\section{Bound on the Rényi entanglement asymmetry for finite discrete groups}\label{appendix:bound}
In this Appendix, we prove that, for a finite discrete group $G$, the Rényi entanglement asymmetry $\Delta S_A^{(n)}$ is bounded by the dimension $N$ of the group. Let us write the projected reduced density matrix $\rho_{A, Q_G}$ of a finite group $G$ of elements $g_1, \dots, g_N$ as
\begin{equation}
    \rho_{A, Q_G} = \frac{1}{N}\sum_{j=1}^N \rho_{j} \text{ where } \rho_{j} = g_j \rho_A g_j^{-1}.
\end{equation}
The trace of $\rho_{A, Q_G}^n$ is given by
\begin{equation}
    \Tr(\rho_{A, Q_G}^n) = \frac{1}{N^n}\sum_{j_1, \cdots, j_n =1}^N \Tr(\rho_{j_1}\cdots \rho_{j_n}).
\end{equation}
Since the $\rho_j$ are all positive semi-definite, 
\begin{equation}
    {\Re}\left[\Tr(\rho_{j_1}\cdots \rho_{j_n})\right]\ge 0, \quad \sum_{j_1, \cdots, j_n =1}^N {\rm Im}\left[\Tr(\rho_{j_1}\cdots \rho_{j_n})\right]=0,
\end{equation}
and they have the same spectrum $\Tr(\rho_{j}^n)=\Tr(\rho_A^n)$, one consequently has
\begin{equation}\label{eq:trace_prod}
    \sum_{j_1, \cdots, j_n =1}^N \Tr(\rho_{j_1}\cdots \rho_{j_n})\ge N \Tr(\rho_A^n).
\end{equation}
Applying this result in the definition~\eqref{eq:def_ren_asym} of the Rényi entanglement asymmetry, one finds that
\begin{equation}\label{eq:renyi_ent_asymm_bound}
    \Delta S_A^{(n)}\le \log N.
\end{equation}
If the state $\rho_A$ maximally breaks the symmetry, i.e. the $\rho_j$ have orthogonal support,
\begin{equation}
    \Tr(\rho_{j} \rho_{j'}) =\delta_{jj'} \Tr(\rho_A^2),
\end{equation}
then if two indices differ in the product~\eqref{eq:trace_prod} of $\rho_{j}$, by the cyclic invariance of the trace we can always write using the Cauchy-Schwarz inequality,
\begin{equation}
    0\le \Tr(\rho_{j_1}\cdots \rho_{j_n}) \le \Tr(\rho_{j_1} \cdots \rho_{j_{n-2}})\Tr(\rho_{j_{n-1}}\rho_{j_n})=0, \quad j_{n-1}\neq j_n,
\end{equation}
and the bound of Eq.~\eqref{eq:renyi_ent_asymm_bound} is then saturated. 

\section{Diagonalisation of the XY spin chain Hamiltonian with OBC}\label{app:diag_obc}

As we discuss in Sec.~\ref{sec:ent_asymm_obc}, in order to diagonalise the Hamiltonian~\eqref{eq:HXY} for Open Boundary Conditions, it is sufficient to know the rotation matrix $\mathcal{R}_{h,\gamma}$ between the Majorana and Bogoliubov fermions introduced in Eq.~\eqref{eq:bogoliubov_rot}. In the quantum Ising line, $\gamma=1$, we obtain it following the detailed discussion presented in Ref.~\cite{kormos}. First, it is necessary to solve the quantisation condition
\begin{equation}\label{eq:quant_cond}
    k = \frac{\pi}{L+1}\left(j+\frac{1}{\pi} \arctan(\frac{\sin k}{\cos k-h})\right)\text{ with }k \in (0, \pi) \text{ and } j\in \{ 1, 2, \dots, L \},
\end{equation}
which has $L-1$ real solutions for $h<1$ and $L$ solutions for $h\ge 1$. For $h<1$ the last mode is the boundary mode, given by the complex momentum $k_0 = \pi + i\nu$ where $\nu$ is obtained by solving
\begin{equation}
    \tanh (\nu (L+1))=\frac{\sinh \nu}{h + \cosh \nu}\quad \text{ with }\nu>0.
\end{equation}
Once we have numerically obtained the momenta with Eq.~\eqref{eq:quant_cond}, we can construct the modes $\eta_k$ that diagonalise the Hamiltonian~\eqref{eq:kitaev_chain} as
\begin{align}
    \eta_{k}^\dagger &= \frac{1}{2}\sum_{j=1}^{2L} A_k \left(\sin(kj-\theta_k) \check a_{2j-1} +i\sin(kj) \check a_{2j} \right),\\
    \eta_{k} &= \frac{1}{2}\sum_{j=1}^{2L} A_k \left(\sin(kj-\theta_k) \check a_{2j-1} -i\sin(kj) \check a_{2j} \right),
\end{align}
where $A_k$ is a normalisation constant,
\begin{equation}
    A_k = \left(\frac{L}{2} - \frac{h(\cos k-h)}{2\epsilon_k^2}\right)^{-\frac 1 2}.
\end{equation}
Similarly, the boundary modes are given by
\begin{align}
    \eta_{k_0}^\dagger = \frac{1}{2}\sum_{j=1}^{2L} A_{k_0} (-1)^j \left(\sinh(\nu (L+1-j)) \check a_{2j-1} +i\sinh(\nu j) \check a_{2j} \right),\\
    \eta_{k_0} = \frac{1}{2}\sum_{j=1}^{2L} A_{k_0} (-1)^j \left(\sinh(\nu (L+1-j)) \check a_{2j-1} -i\sinh(\nu j) \check a_{2j} \right),
\end{align}
where 
\begin{equation}
    A_{k_0} = \left(\sum_{j=1}^{L}\sinh^2(\nu (L+1-j))+\sinh^2(\nu j) \right)^{-\frac 1 2}.
\end{equation}

\section{Gaussian density matrices and their composition rules}\label{app:gaussian_rules}
In general, a Gaussian density matrix is generated from a quadratic form such that
\begin{equation}
    \rho_W= \frac{1}{\mathcal Z(W)} \exp(\frac{1}{4}\sum_{m,m'}W_{mm'} \check a_m \check a_{m'}),
\end{equation}
where the factor $\mathcal{Z}(W)$ ensures that $\Tr(\rho_W)=1$. A Gaussian density matrix $\rho_W$ satisfies the Wick theorem and, therefore, it is univocally determined by the two-point correlation matrix~\cite{peschel}
\begin{equation}
    \delta_{mm'} + i\Gamma_{mm'}= \Tr(\rho_W \check a_m \check a_{m'}),
\end{equation}
which can be written as
\begin{equation}
\Gamma=i\tanh\left(\frac{W}{2}\right).
\end{equation}
We can thus use the notation $\rho_W\equiv \rho[\Gamma]$. For large subsystems, it is in practice impossible to directly use full reduced density matrices because their dimension grows exponentially as $2^\ell$ with the subsystem size $\ell$, whereas correlation matrices have a size $2\ell\times 2\ell$.

The computation of the Rényi entanglement asymmetry~\eqref{eq:def_ren_asym} involves traces of products of different Gaussian matrices $\Tr\left(\rho[\Gamma_1] \cdots \rho[\Gamma_n]\right)$. To that end, let us first consider two Gaussian density matrices $\rho[\Gamma]$ and $\rho[\Gamma']$ associated to the correlation matrices $\Gamma$ and $\Gamma'$. The special properties of Gaussian operators allow to write their product in the form\cite{balian, fc-10}
\begin{equation}
    \rho[\Gamma] \rho[\Gamma'] = \Tr[\rho[\Gamma] \rho[\Gamma']] \rho[\Gamma \times \Gamma'],
\end{equation}
where the composition rule $\times$ for correlation matrices is defined as
\begin{equation}
    \Gamma \times \Gamma' := i\left[I - (I+ i \Gamma')(I - \Gamma \Gamma')^{-1} (I + i \Gamma) \right]
\end{equation}
and the trace of two Gaussian density matrices can be obtained from their respective correlation matrices as
\begin{equation}
    \{\Gamma, \Gamma'\}:=\Tr\left(\rho[\Gamma]\rho[\Gamma']\right)  = \prod_{j} \frac{1-\mu_j}{2} = \pm \sqrt{\det (\frac{I - \Gamma \Gamma'}{2})}
\end{equation}
where $\mu_j$ are the eigenvalues of $\Gamma \Gamma'$ with halved degeneracy. Observe that, in principle, there is an ambiguity in the global sign in the last equality of the expression above, which is fixed in our case since the traces we compute always have a positive sign.

Using these rules, any trace of a product of Gaussian operators can be expressed as
\begin{equation}\label{eq:trace_gaussians}
    \Tr[\rho_1 \cdots \rho_n] = \{\Gamma_1, \cdots, \Gamma_n\},
\end{equation}
where 
\begin{equation}
    \{\Gamma_1, \cdots, \Gamma_n\} = \{\Gamma_1, \Gamma_2\} \{\Gamma_1 \times \Gamma_2, \Gamma_3, \cdots, \Gamma_n\}
\end{equation}
can be computed by recurrence.

\section{Derivation of the expressions to calculate $\Delta S_A^{(n)}$}
In this Appendix, we describe in detail how we obtain the formulas of Eqs.~\eqref{eq:ent_asymm_obc} and \eqref{eq:ent_asymm_pbc} that we use in the main text to compute efficiently the time evolution of entanglement asymmetry after a quench in OBC and PBC respectively.

\subsection{In OBC}\label{app:ent_asymm_obc}

According to Eq.~\eqref{eq:renyi_asymm_even_op}, in order to calculate the Rényi entanglement asymmetry we must determine the traces $\Tr(\rho_{A, e}^n)$ and $\Tr(\rho_{A}^n|_{e})$. In Sec.~\ref{sec:obc}, we have seen that in OBC the reduced density matrix $\rho_A$ is given by the ansatz of Eq.~\eqref{eq:ansatz}. Therefore, in this case, the even and odd parts of $\rho_A$ are
\begin{equation}\label{eq:even_odd_rhoa_obc}
    \rho_{A, e} = \rho_0, \quad \rho_{A, o} = \frac 1 2 \langle \check b_A \rangle (\check b_A \rho_0 + \rho_0 \check b_A ).
\end{equation}
The $n$th power of $\rho_A$ can be then obtained as
\begin{equation}\label{eq:rhoan}
    \rho_A^n = \sum_{\{\pi_j=e, o\}}\prod_{j=1}^n \rho_{A, \pi_j}.
\end{equation}
Therefore, its even part $\rho_A^n|_e$ are the terms in this sum that include an even number of times the factor $\rho_{A, o}$.

Let us now introduce the variable $\chi_j$ corresponding to the parity of each factor, i.e. $\chi_j=0$ for $\pi_j=e$ and $\chi_j=1$ for $\pi_j=o$, such that the even and odd parts~\eqref{eq:even_odd_rhoa_obc} of $\rho_A$ can be expressed in the same foot,
\begin{equation}
    \rho_{A, \pi_j} = \frac{1}{2} \langle \check b_A \rangle^{\chi_j} \check b_A^{\chi_j} (\rho_0 + \check b_A^{\chi_j} \rho_0 \check b_A^{\chi_j}),
\end{equation}
where
\begin{equation}
    \check b_A^0 \equiv I, \quad \check b_A^1 \equiv \check b_A.
\end{equation}
Observe that, in a product of several matrices $\rho_{A, \pi_j}$, each factor can be written as $(\rho_0 + \check b_A^{\chi_j} \rho_0 \check b_A^{\chi_j})$ with an extra Majorana fermion $\langle \check b_A\rangle \check b_A$ inserted on the left if it corresponds to $\rho_{A, o}$. Using now the fact that $\check b_A^2 \equiv I$, we can rewrite the product e.g. 
\begin{equation}
    \rho_{A, o} \rho_{A, e} \rho_{A, o} =\langle \check b_A\rangle^2 \check b_A( \frac 1 2 \rho_0 +\frac 1 2 \check b_A \rho_0 \check b_A)\rho_0 \check b_A (\frac 1 2 \rho_0 + \frac 1 2 \check b_A \rho_0 \check b_A)
\end{equation}
in the form
\begin{equation}
   \rho_{A, o} \rho_{A, e} \rho_{A, o} =\langle \check b_A\rangle^2 \left[\check b_A( \frac 1 2 \rho_0 +\frac 1 2 \check b_A \rho_0 \check b_A) \check b_A\right] \left[\check b_A\rho_0 \check b_A\right] (\frac 1 2 \rho_0 + \frac 1 2 \check b_A \rho_0 \check b_A).
\end{equation}
That is, whenever we have a product of the matrices $\rho_{A, e}$ and $\rho_{A, o}$ that contains an even number of terms $\rho_{A, o}$, we can rewrite it as
\begin{equation}\label{eq:rhonobc0}
\prod_{j=1}^n \rho_{A, \pi_j}=\frac{1}{2^n}\prod_{j=1}^n \check b_A^{s_j} (\rho_0 + \check b_A^{\chi_j} \rho_0 \check b_A^{\chi_j}) b_A^{s_j},
\end{equation}
where $s_j=\sum_{j'=1}^j \chi_j$. Observe that we insert the operator $\check b_A$ on the left and the right of the $j$ factor $(\rho_0 + \check b_A^{\chi_j} \rho_0 \check b_A^{\chi_j})$ if until that position there is an odd number of terms $\rho_{A,o}$. Consequently, it is possible to write the even part of $\rho_A^n$ as
\begin{equation}\label{eq:rhonobc_1}
    \rho_A^n|_e = \frac{1}{2^n} \sum_{\{\chi_i=0, 1\}} \frac{1+(-1)^{s_n}}{2} \langle \check b_A \rangle^{s_n} \prod_{j=1}^n \check b^{s_j}(\rho_0 + \check b^{\chi_j} \rho_{0} \check b^{\chi_j}) \check b^{s_j},
\end{equation}
where the coefficient $(1+(-1)^{s_n})/2$ ensures that we only take products containing an even number of $\rho_{A, o}$. Since $\check b_A^2\equiv I$, it is now useful to perform in Eq.~\eqref{eq:rhonpbc_1} the change of variables $w_i = s_i (\mod 2)$, 
\begin{equation}\label{eq:rhonobc_2}
    \rho_A^n|_e = \frac{1}{2^n} \sum_{\substack{\{w_i=0, 1\} \\ w_0=w_n=0}}  \prod_{j=1}^n \langle \check b_A \rangle^{|w_j-w_{j-1}|}( \check b_A^{w_j}\rho_0 \check b_A^{w_j} + \check b_A^{w_{j-1}} \rho_{0} \check b_A^{w_{j-1}}),
\end{equation}
where the condition $w_0=0$ is due to the fact that the empty sum is $0$ and $w_n = 0$ ensures that we take an even number of factors $\rho_{A, o}$. Eq.~\eqref{eq:rhonobc_2} can be rewritten as
\begin{equation}\label{eq:rhonobc_3}
    \rho_A^n|_e = \frac{1}{2^n} \sum_{\substack{\{w_i=0, 1\} \\ w_0=w_n=0}}  \prod_{j=1}^n \langle \check b_A \rangle^{|w_j-w_{j-1}|}\left((w_j+w_{j-1})\rho_0 + (2-w_j-w_{j-1}) \check b_A\rho_0 \check b_A\right).
\end{equation}
Once expanded, it finally takes the form
\begin{equation}\label{eq:rhonpbc_3}
\rho_A^n|_e=\frac{1}{2^n}\sum_{\{\tau_i=0, 1\}}\zeta[\{\tau_i\}]\prod_{j=1}^n \check b_A^{\tau_j}\rho_0 \check b_A^{\tau_j},
\end{equation}
where 
\begin{equation}
    \zeta [\{\tau_i\}] = \sum_{\substack{\{w_i=0, 1\} \\ w_0=w_n=0}}\prod_{j=1}^n \Biggl(\langle\check b_A\rangle^{|w_j -w_{j-1}|} \Bigl(1 + (-1)^{\tau_j}(1-w_j-w_{j-1})\Bigl) \Biggl).
\end{equation}
According to Eq.~\eqref{eq:rhonobc_3}, the calculation of the trace $\Tr(\rho_A^n|_e)$ boils down to compute the traces of products
$\Tr(\prod_{j=1} \check b_A^{\chi_j} \rho_0 b_a^{\chi_j})$. One can show that the matrix $\check b_A \rho_0 \check b_A$ is Gaussian,
\begin{equation}
    \check b_A \rho[\Gamma_0^A] \check b_A = \rho[\Gamma_1^A],
\end{equation}
and is univocally determined by the correlation matrix $(\Gamma_1^A)_{mm'}=\Tr(\check b_A \rho_0 \check b_A a_m a_{m'})$, $m, m'\in A$, whose entries are given by the expression reported in Eq.~\eqref{eq:gamma_1_obc}. Therefore, we can apply the formula of Eq.~\eqref{eq:trace_gaussians} to compute the trace of a product of Gaussian operators in terms of their correlation matrices. We then find
\begin{equation}\label{eq:trace_rhoa_n_even_obc}
\Tr(\rho_A^n|_e)=\frac{1}{2^n}\sum_{\{\tau_i=0, 1\}}\zeta[\{\tau_i\}]\{\Gamma_{\tau_1}^A, \cdots, \Gamma_{\tau_n}^A\}.
\end{equation}

The other ingredient left is the trace $\Tr(\rho_{A, e}^n)$, since $\rho_{A, e}=\rho_0$ is a Gaussian operator with correlation matrix 
$\Gamma_0^A$, then the straightforward application of Eq.~\eqref{eq:trace_gaussians} gives
\begin{equation}\label{eq:trace_rhoa_even_n_obc}
\Tr(\rho_{A, e}^n)=\{\Gamma_0^A, \dots, \Gamma_0^A\}.
\end{equation}
Plugging the results of Eqs.~\eqref{eq:trace_rhoa_n_even_obc} and \eqref{eq:trace_rhoa_even_n_obc} in Eq.~\eqref{eq:renyi_asymm_even_op}, we obtain the expression~\eqref{eq:ent_asymm_obc}.

\subsection{In PBC}\label{app:ent_asymm_pbc}

In PBC, the reduced density matrix $\rho_A$ takes the form of Eq.~\eqref{eq:rdmpbc}, and its even and odd parts correspond to
\begin{equation}
\rho_{A, e}=\rho_0, \quad  \rho_{A, o}=\langle \sigma_1^x\rangle \sigma_1^x \rho_1.
\end{equation}
As in OBC, the $n$th power of $\rho_A$ takes the form of 
Eq.~\eqref{eq:rhoan}. To compute the entanglement asymmetry~\eqref{eq:renyi_asymm_even_op}, we 
need to obtain the trace of the even part of $\rho_A^n$, which 
consists of the terms of the sum~\eqref{eq:rhoan} that contain an 
even number of factors $\rho_{A, o}$. 

Let us consider a particular product of operators $\rho_{A, e}$ and $\rho_{A, o}$, similar to the ones that appear in Eq.~\eqref{eq:rhoan}. For example,
\begin{equation}\label{eq:rdm_string_2}
    \rho_{A, e} \rho_{A, o} \rho_{A, e} \rho_{A, e} \rho_{A, o} = \langle \sigma^x_1 \rangle^2 \rho_{0} \sigma^x_1 \rho_1 \rho_0 \rho_0 \sigma^x_1 \rho_1.
\end{equation}
We can now use the fact that $\sigma^x_1 = \check a_1$, and that the product at both sides of a Gaussian operator by a Majorana fermionic excitation, e.g. $\check a_1 \rho_0 \check a_1$ and $\check a_1 \rho_1 \check a_1$, is also Gaussian. Since $\check a_1^2 = 1$, Eq.~\eqref{eq:rdm_string_2} can be then re-expressed as the product of Gaussian operators
\begin{equation}
    \rho_{A, e} \rho_{A, o} \rho_{A, e} \rho_{A, e} \rho_{A, o} = \langle \sigma^x_1 \rangle^2 \rho_0 (\check a_1 \rho_1 \check a_1)( \check a_1 \rho_0 \check a_1)( \check a_1 \rho_0 \check a_1) \rho_1.
\end{equation}
Observe that, whether or not the Majorana fermion $\check a_1$ multiplies at both sides either $\rho_0$ or $\rho_1$ at position $j$ depends on the number of $\rho_{A, o}$ factors in the product up to that position. If the number of $\rho_{A, o}$ is odd in the total string, there is an additional $\check a_1$ term at the end and the trace vanishes.

Using the previous observation, we can write the even part of $\rho_A^n$ as
\begin{equation}\label{eq:rhonpbc_1}
    \rho_A^n|_e = \sum_{\{\chi_i=0, 1\}} \frac{1+(-1)^{s_n}}{2} \langle \sigma^x_1 \rangle^{s_n} \prod_{j=1}^n (\check a_1^{s_j} \rho_{\chi_j} \check a_1^{s_j}),
\end{equation}
where we defined $s_j = \sum_{i=1}^j \chi_i$ and the term $(1+(-1)^{s_n})/2$ ensures that we only take products containing an even number of $\rho_{A, o}$. Thanks to Eq.~\eqref{eq:rhonpbc_1}, the trace of $\rho_A^n|_e$ can be obtained as a sum of traces of products of Gaussian operators 
$\Tr\left(\prod_{j=1}^n\check a_1^{s_j}\rho_{\chi_j}\check a_1^{s_j}\right)$, which can be computed in terms of their correlation matrices using Eq.~\eqref{eq:trace_gaussians}.
The only missing step is thus to determine the correlation matrices of $\check a_1 \rho_0 \check a_1$ and $\check a_1 \rho_1 \check a_1$. Employing the anticommutation relations of Majorana fermions, we can obtain the identity
\begin{equation}
    \check a_1 \rho[\Gamma]\check a_1 = \rho[P_1 \Gamma P_1]
\end{equation}
for any Gaussian matrix $\rho[\Gamma]$, where $P_1 = {\rm diag}(-1, 1, \cdots, 1)$. Applying this equality, we conclude that the Gaussian operator $\check a_1 \rho_{\chi_j} \check a_1$ is determined by the correlation matrix $P_1 \Gamma_{\chi_j}^A P_1$.  Therefore, using  the identity~\eqref{eq:trace_gaussians} when we take the trace in Eq.~\eqref{eq:rhonpbc_1}, we have
\begin{equation}\label{eq:trace_rhoa_n_even_pbc}
\Tr(\rho_{A}^n|_e)= \sum_{\{\chi_i=0, 1\}}\frac{1+(-1)^{s_n}}{2}\langle \sigma_1^x\rangle^{s_n}\{P_1^{s_1}\Gamma_{\chi_1}^AP_1^{s_1}, \cdots, P_{1}^{s_n}\Gamma_{\chi_i}^AP_{1}^{s_n}\}.
\end{equation}
On the other hand, since $\rho_{A, e}=\rho_0$, we can directly apply Eq.~\eqref{eq:trace_gaussians} to calculate $\Tr(\rho_{A, e}^n)$ in terms 
of the correlation matrix $\Gamma_0^A$,
\begin{equation}\label{eq:trace_rhoa_even_n_pbc}
    \Tr(\rho_{A, e}^n) = \{\Gamma_0^A, \overset{n}{\dots}, \Gamma_0^A\}.
\end{equation}
Plugging Eqs.~\eqref{eq:trace_rhoa_n_even_pbc} and \eqref{eq:trace_rhoa_even_n_pbc} in Eq.~\eqref{eq:renyi_asymm_even_op}, we finally find the formula~\eqref{eq:ent_asymm_pbc} to study the entanglement asymmetry of spin parity in PBC.

\section{Obtaining $\rho_{A, {o}}$ with cluster decomposition}\label{appendix:cluster_decomposition}

In this Appendix, we show how to write the reduced density matrix $\rho_A$ for the ground state~\eqref{eq:pbcstate} in the form of Eq.~\eqref{eq:rdmpbc}. Our goal is to reconstruct the odd part of it, $\rho_{A, {o}}$, in terms of a Gaussian operator. Since we are in PBC, without loss of generality and to simplify the notation, we will take as subsystem $A$ an interval from the sites $j=1$ to $j=\ell$, that is, attached to the beginning of the Jordan-Wigner string~\eqref{eq:jordanwigner}. First, let us take the density matrix of the total system in the ground state~\eqref{eq:pbcstate}, $\rho=\ket{\Psi}\bra{\Psi}$. It can be decomposed in even and odd parts
\begin{equation}
\rho=\rho_{e}+\rho_{o},
\end{equation}
where
\begin{equation}
\rho_{e}=\frac{1}{2}\ket{\emptyset}_{NS}\prescript{}{NS}{\bra{\emptyset}} + \frac{1}{2}\ket{\emptyset}_{R}\prescript{}{R}{\bra{\emptyset}}
\end{equation}
and 
\begin{equation}
\rho_{o}=\frac{1}{2}\ket{\emptyset}_{R}\prescript{}{NS}{\bra{\emptyset}}+ \frac{1}{2}\ket{\emptyset}_{NS}\prescript{}{R}{\bra{\emptyset}}.
\end{equation}
In principle, $\rho_{A, o}=\Tr_B(\rho_o)$ but this is an odd operator while Gaussian ones are even.  Inspired by Ref.~\cite{fc-10}, the strategy is to construct a fake density matrix $\rho_1$ multiplying $\rho_{A, o}$  by an odd operator, such as $\sigma_1^x$, 
\begin{equation}
\rho_1=\frac{\sigma_1^x}{\langle \sigma_1^x\rangle } \rho_{A,o},
\end{equation}
where the factor $1/\langle \sigma_1^x\rangle$ ensures that $\Tr \rho_1=1$. Observe that  $\rho_1=\Tr_B(\sigma_1^x \rho_{o})/\langle \sigma_1^x\rangle$ since for any operator $\mathcal{O}_A$ with support on $A$,
\begin{equation}
\Tr_A(\Tr_B(\sigma_1^x \rho_{o})\mathcal{O}_A)=\Tr(\sigma_1^x \rho_{o} \mathcal{O_A})=\Tr_A(\sigma_1^x \rho_{A, o} \mathcal{O}_A) 
\end{equation}
and $\sigma_1^x$ acts on $\mathcal{H}_A$.

\begin{figure}[t]
  \centering
  \includegraphics[width=.8\textwidth]{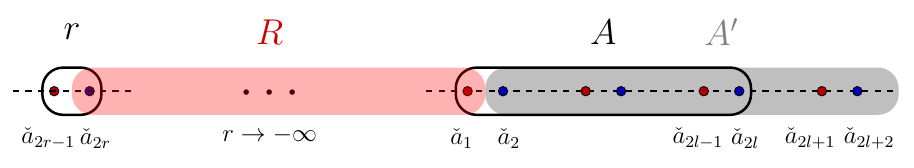}
  \caption{Setup for the cluster decomposition described in the text to determine the operator $\rho_1$.}
  \label{fig:cluster_dec_4}
\end{figure}

Although $\rho_1$ is an even operator as such it can not be computed in any practical way. This is where cluster decomposition enters in the game. We consider an auxiliary spin $\sigma_r^x$ at the site $r$ and let us introduce 
\begin{equation}
\rho_{1, r}=\frac{\Tr_B(\sigma_1^x \sigma_r^x \rho_{e})}{\langle \sigma_1^x \sigma_r^x \rangle}.
\end{equation}
If we consider an arbitrary operator $\mathcal{O}_A$ with support in $A$, then
\begin{equation}
\Tr_{A}(\rho_{1, r}\mathcal{O}_A)=\frac{\prescript{}{NS}{\bra{\emptyset}} \mathcal{O}_A \sigma_1^x \sigma_r^x \ket{\emptyset}_{NS}}{\langle \sigma_1^x \sigma_r^x \rangle} = \begin{cases}
    0 \text{ if $\mathcal{O}_A$ is odd,}\\[10pt]
    \dfrac{\langle \mathcal{O}_A \sigma_1^x \sigma_r^x \rangle}{\langle \sigma_1^x \sigma_r^x \rangle}\text{ if $\mathcal{O}_A$ is even.}
\end{cases}
\end{equation}
If we now take the limit $r\to \pm \infty$ in the expression above, we can use the cluster decomposition principle so that
\begin{equation}
\lim_{r\to \pm \infty} \Tr_A(\rho_{1, r} \mathcal{O}_A)= \left.\begin{cases}
    0 \text{ if $\mathcal{O}_A$ is odd}\\[10pt]
    \dfrac{\langle \mathcal{O}_A \sigma_1^x \rangle}{\langle \sigma_1^x \rangle}\text{ if $\mathcal{O}_A$ is even}
\end{cases}\right\} =\Tr(\rho_1 \mathcal{O}_A).
\end{equation}
Therefore, $\rho_{1, r}$ is Gaussian since it is built  from the even
part of $\rho$, which is Gaussian itself and, consequently, it is univocally determined by a two-point correlation matrix. Moreover, it correctly reproduces the same correlations as $\rho_{1}$ when $r\to \pm \infty$. For practical reasons, we will now take $r\to -\infty$ so that the Jordan-Wigner string between the sites $1$ and $r$ does not belong to the subsystem $A$. We can now write
\begin{equation}
    \rho_1 = \lim_{r\to - \infty} \rho_{1, r}.
\end{equation}
 The correlation matrix $\Gamma_{1, r}^A$ that determines $\rho_{1, r}$ has entries $\delta_{mm'}+i(\Gamma_{1, r}^A)_{mm'}=\langle \check a_m \check a_{m'}\sigma_1^x \sigma_r^x \rangle$, with $m, m'\in A$, which are expectation values of strings of Majorana fermions since  
\begin{equation}\label{eq:stringPtilde}
    P_R:=\sigma^x_1 \sigma^x_r = \sigma^x_r \sigma^x_1=\prod_{j=r}^0 (-i\check a_{2j} \check a_{2j+1}).
\end{equation}
Consequently the correlation matrix of $\rho_{1,r}$ has the form
\begin{equation}\label{eq:stringcov}
    (\delta_{mm'}+i\Gamma_{1, r}^A)_{mm'} = \frac{\langle \check a_m \check a_{m'} P_R \rangle}{\langle P_R \rangle}, \text{ with }m, m'\in A.
\end{equation}
These correlations can be computed as a Pfaffian of a submatrix of $\Gamma_0$, defined in Eq.~\eqref{eq:cor1pbc}, as shown in Ref.~\cite{cef-12-1}. However, a more computationally efficient way to obtain them is to use the results on reduced density matrices of disjoint intervals found in Ref.~\cite{fc-10}. Indeed, correlation matrices of the form \eqref{eq:stringcov} can be computed as Schur complements. Since the string of Eq.~\eqref{eq:stringPtilde} overlaps with subsystem $A$ at $m=1$, we need to consider the subsystem $A'$ as defined in Fig.~\ref{fig:cluster_dec_4}. Then we introduce the correlation matrix $\Gamma_{1, r}^{A'}$ with entries $\delta_{mm'}+i(\Gamma_{1, r}^{A'})_{mm'}=\langle \check a_m \check a_{m'} P_R\rangle$, with $m, m'\in A'$.
\paragraph{}
Let us now consider the restriction of the two-point correlation matrix $\Gamma_{0}$ to the subsystem $R\cup A'$,
\[\Gamma_{0}^{R\cup A'} = \begin{pNiceMatrix}[columns-width=0.3cm]
 \Block[draw,fill=red!25,rounded-corners]{2-2}{\Gamma_{0}^R} & &\Block{2-2}{\Gamma_0^{R,A'}} \\
 &  & &  \\
\Block{2-2}{\Gamma_0^{A',R}} &  &  \Block[draw,fill=gray!25,rounded-corners]{2-2}{\Gamma_0^{A'}} &  \\
 & & &
\end{pNiceMatrix}.\]
The correlation matrix $\Gamma_{1, r}^{A'}$ can then be obtained as a Schur complement~\cite{fc-10}
\begin{equation}
\Gamma_{1, r}^{A'} = \Gamma_0^{A'} - \Gamma_0^{A', R}(\Gamma_0^{R})^{-1}\Gamma_0^{R, A'}
\end{equation}
and it has the following structure:
\NiceMatrixOptions%
{code-for-first-row = \scriptstyle}
\[
\Gamma_{1,r}^{A'} = \begin{pNiceMatrix}[columns-width=auto, first-row]
\check a_2 & \check a_3  & \check a_4 & \Cdots & \check a_{2l}& \check a_{2l+1} & \check a_{2l+2}\\
 
 \Block[draw, color=green,rounded-corners]{5-5}{\Gamma_{1, r}^{A\cap A'}} & &\Block[draw, color=blue, rounded-corners]{1-5}{\Gamma_{1, r}^{2,j+2}} & & & & \\
 
  & & & & & & \\
 
 \Block[draw, color=blue, rounded-corners]{5-1}{\Gamma_{1, r}^{j+2,2}} & & & & & & \\
   &  & & & & &  \\
   &  & & & & & \\
  & & & & & & \\
  & & & & & & \\
\end{pNiceMatrix}.\]
By definition $(\Gamma_{1, r}^{A'})_{mn} = (\Gamma_1^A)_{mn}$ for the sites $m, n\in A\cap A'$. Using now the translational invariance of the chain and the fact that when $r\to -\infty$ we can take a separation of one less site with no consequence, we have
\begin{equation}
    \langle \check a_1 \check a_j P_R \rangle = \langle \check a_1 \check a_j (-i\check a_0 \check a_1) \prod_{j=r}^{-1} (-i \check a_{2j} \check a_{2j+1}) \rangle = i\langle \check a_0 \check a_j \prod_{j=r}^{-1} (-i \check a_{2j} \check a_{2j+1}) \rangle \underset{r\to \infty}{=} i\langle \check a_2 \check a_{j+2} P_R \rangle.
\end{equation}
Therefore, we can finally reconstruct $\Gamma_1^A$ as
\[
\Gamma_1^A = \begin{pNiceMatrix}[columns-width=auto, first-row]
\check a_1 & \Cdots &  & \check a_{2l} \\
    0 & \Block[draw, color=blue, rounded-corners]{1-3}{i \Gamma_{1, r}^{2,j+2}} & &\\
    \Block[draw, color=blue, rounded-corners]{3-1}{i\Gamma_{1, r}^{j+2,2}} & \Block[draw, color=green,rounded-corners]{3-3}{\Gamma_{1, r}^{A\cap A'}} & &\\
    & & & \\
    & & & \\
\end{pNiceMatrix}.
\]
Note that all this construction is based on the assumption that $\langle \sigma_1^x \rangle \neq 0$. It is a priori possible to derive $\rho_{A, o}$ from other odd correlators, but we found that it is always sufficient numerically to use this expression. When $\langle \sigma^x_1\rangle$ vanishes because of oscillations after the quench, we simply used the same reasoning applied to $\sigma^y_1$ and $\sigma^y_r$ to cure the divergence.

\end{document}